\documentclass[twocolumn,showpacs,preprintnumbers,,nofootinbib,eqsecnum,amsmath,amssymb]{revtex4}
\usepackage{graphicx}
\usepackage{bm}
\usepackage{epsfig}
\usepackage{amsmath}
\usepackage{amsfonts}
\usepackage{amssymb}%
\usepackage{dcolumn}% Align table columns on decimal point
\begin{document}
%%%%%%%%%%%%%%%DEFINTIONS%%%%%%%%%%%%%%%
\def\gsim{\:\raisebox{-0.5ex}{$\stackrel{\textstyle>}{\sim}$}\:}
\def\etal{{\it et al.}}
\def\go{\rightarrow  }
\def\be{\begin{equation}}
\def\ee{\end{equation}}
\def\br{\begin{eqnarray}}
\def\er{\end{eqnarray}}
\def\brn{\begin{eqnarray*}}
\def\ern{\end{eqnarray*}}
\def\rf#1{{(\ref{#1})}}
\def\a {{\alpha}}
\def\b {{\beta}}
\def\e {{\epsilon}}
\def\k {{\kappa}}
\def\l {{\lambda}}
\def\s {{\sigma}}
\def\w {{\omega}}
\def\sss{\scriptscriptstyle}
\def\nn{\nonumber}
\def\ie{{\em i.e., }}
\def\x{\times}
\def\F {{{\cal F}}}
\def\L {{{\cal L}}}
\def\M {{{\cal M}}}
\def\T {{{\cal T}}}
\def\pb {{\bf p}}
\def\qb {{\bf q}}
\def\q {{\rm q}}
\def\d{\dagger}
\def\ket#1{|#1 \rangle}
\def\bra#1{\langle #1|}
\def\Ket#1{||#1 \rangle}
\def\Bra#1{\langle #1||}
\def\ss{\scriptstyle}
\def\ol{\overline}
\def\etc{ {\it etc}}
\def\up{u_{{\rm p}}}
\def\vp{v_{{\rm p}}}
\def\un{u_{{\rm n}}}
\def\vn{v_{{\rm n}}}
\def\upp{u_{p'}}
\def\vpp{v_{p'}}
\def\unp{u_{n'}}
\def\vnp{v_{n'}}
\def\vk{v_{{\rm k}}}
\def\uk{u_{{\rm k}}}
\def\del {\delta}
\def\sl{\textsl{}}
\def\sss{\scriptscriptstyle}
\def\kb {{\bf k}}
\def\rb {{\bf r}}
\def\vb {{\bf v}}
\def\xb {{\bf x}}
\def\sq{\sqrt{2}}
\def\g {{\gamma}}
\def\gAeff{g_{\mbox{\tiny A,eff}}}
\def\gV{g_{\mbox{\tiny V}}}
\def\gM{g_{\mbox{\tiny M}}}
\def\gS{g_{\mbox{\tiny S}}}
\def\gT{g_{\mbox{\tiny T}}}
\def\gP{g_{\mbox{\tiny P}}}
\newcommand{\Mass}{\mathrm{M}}
\def\gA{g_{\mbox{\tiny A}}}
\newcommand{\mass}{\mathrm{m}}
\def\Jb {{\bf J}}
\def\lb {{\bf l}}
\def\ga{\overline{g}_{\mbox{\tiny A}}}
\def\gv{\overline{g}_{\mbox{\tiny V}}}
\def\gp{\overline{g}_{\mbox{\tiny P}}}
\def\gpa{\overline{g}_{\mbox{\tiny P1}}}
\def\gpb{\overline{g}_{\mbox{\tiny P2}}}
\def\gw{\overline{g}_{\mbox{\tiny W}}}
\def\gm{\overline{g}_{\mbox{\tiny M}}}
\def\mbs{\mbox{\boldmath$\sigma$}}
\def\mbn{\mbox{\boldmath$\nabla$}}
\def\Ob {{\bf O}}
\def\kr {{\bf k}\cdot{\bf r}}
\def\qr {{\bf q}\cdot{\bf r}}
\def\threej#1#2#3#4#5#6{\left(\negthinspace\begin{array}{ccc}
#1&#2&#3\\#4&#5&#6\end{array}\right)}
\def\sixj#1#2#3#4#5#6{\left\{\negthinspace\begin{array}{ccc}
#1&#2&#3\\#4&#5&#6\end{array}\right\}}
\def\ninej#1#2#3#4#5#6#7#8#9{\left\{\negthinspace\begin{array}{ccc}
#1&#2&#3\\#4&#5&#6\\#7&#8&#9\end{array}\right\}}
\def\fot{\frac{1}{2}}
\def\mbb{\mbox{\boldmath$\beta$}}
\def\absk {{|\kb|}}
\def\etc{ {\it etc}}
\def\bit{\begin{itemize}}
\def\eit{\end{itemize}}
\def\bnu{\begin{enumerate}}
\def\enu{\end{enumerate}}
\def\S {{{\cal S}}}
\def\N {{{\cal N}}}
\def\gvs{g_{\sss{V}}}
\def\gas{g_{\sss{A}}}

%%%%%%%%%%%%%%%%%%%%%%%%%%%%%%%%%%%%%%%%
\title{QRAP: a numerical code for projected (Q)uasi-particle (RA)ndom (P)hase approximation}
\author{A.~R. Samana$^{1,2}$, F. Krmpoti\'c$^{3,4}$ and
C.~A. Bertulani$^{1}$}
\email{krmpotic@fisica.unlp.edu.ar,
arturo_samana@tamu-commerce.edu, carlos_bertulani@tamu-commerce.edu}
\affiliation{$^{1}$Department of Physics, Texas A\&M University
Commerce,  P.O.3011 Commerce, 75429 TX, USA}%
\affiliation{$^2$Departamento de Ci\^encias Exactas e Tecnol\'ogicas,
Universidade Estadual  de Santa Cruz,
CEP 45662-000 Ilheús, Bahia-BA, Brazil}%
\affiliation{$^3$Instituto de F\'{\i}sica La Plata,
CONICET, 1900 La Plata, Argentina}%
\affiliation{$^4$Facultad de Ciencias Astron\'omicas y Geof\'{\i}sicas,
Universidad Nacional de La Plata, 1900 La Plata, Argentina}%
%\affiliation{$^5$Departamento de F\'{\i}sica, Universidad Nacional de
%La Plata, C. C. 67, 1900 La Plata, Argentina}%

\date{\today}

\begin{abstract}

{\bf Abstract}

A computer code for quasiparticle random phase approximation-QRPA
and projected quasiparticle random phase approximation-PQRPA models
of nuclear structure is explained in details. The residual
interaction is approximated by a simple $\delta$-force.
An important application of the code consists in evaluating
nuclear matrix elements involved in neutrino-nucleus reactions.
As an example, cross section for $^{56}$Fe and $^{12}$C are
calculated and the code output is explained. The application
to other nuclei and the description of other nuclear and weak
decay processes is also discussed.

\bigskip\bigskip

{\bf Program summary}

\textit{Title of program:} QRAP ({\bf Q}uasiparticle
{\bf RA}ndom {\bf P}hase approximation)

\textit{Computers:} The code has been created on an PC,
but also runs on UNIX or LINUX machines.

\textit{Operating systems:} WINDOWS or UNIX

\textit{Program language used:} Fortran-77

\textit{Memory required to execute with typical data:}
16 Mbytes of RAM memory and 2 MB of hard disk space

\textit{No. of lines in distributed program, including
test data, etc.: $\sim$ 8,000}

\textit{No. of bytes in distributed program, including
test data, etc.: $\sim$ 256 kB}

\textit{Distribution format:} tar.gz


\textit{Nature of physical problem:} The program calculates neutrino-
and antineutrino-nucleus cross sections as a function of the
incident neutrino energy, and muon capture rates, using the QRPA or
PQRPA as nuclear structure models.

\textit{Method of solution:} The QRPA, or PQRPA, equations are solved
in a self-consistent way for even-even nuclei.
The nuclear matrix elements
for the neutrino-nucleus interaction are treated as the beta inverse
reaction of odd-odd nuclei as function of the transfer momentum.

\textit{Typical running time: $\approx$ 5 min on a 3 GHz processor
for Data set 1.}

\end{abstract}

\maketitle

%\pagebreak

\newpage
{\bf Long Write-Up}

\section{Introduction}

The new age of the physics beyond the standard model of electroweak
interaction has as one of the most promising pathways the search of
neutrino oscillations. Several experimental efforts are oriented to
find the neutrino masses and the related oscillations involving
atmospheric, solar, reactor and accelerator
neutrinos~\cite{Agu01,Fuk98,Aha05,Ara04,Ahn03}. Since neutrinos
interact so weakly with matter, they bring information on the
dynamics of supernova collapse and posterior explosion as well as on
the synthesis of heavy nuclei~\cite{Mcl95,Qia07}.

The detection signal of neutrinos is measured trough the weak
interaction of incoming neutrinos with the nuclei present in, e.g.,
a liquid scintillator detector, as well as with the surrounding
blockhouse detector-shield. The flux-averaged $\nu$-nucleus cross
sections are the measured observables. Recently, Ref. \cite{Aga07}
has studied the effect of neutrino oscillations on the expected
supernova neutrino signal with the LVD detector, through their
interactions with protons and carbon nuclei in a liquid scintillator
and with iron nuclei in the support structure.

Charged and neutral $\nu_e$-nucleus cross sections on $^{12}$C
(liquid scintillator) as well as on $^{56}$Fe (detector surrounding
shield) were measured by the KARMEN
Collaboration~\cite{Mas98,Arm02}. Other experiments such as
LAMPF~\cite{All90,Kra92} and LSND~\cite{Ath96,Ath98} have also used
$^{12}$C to search for neutrino oscillations and to measure
neutrino-nucleus cross sections. Furthermore, future experiments
will use $^{12}$C as liquid scintillator, such as in the spallation
neutron source (SNS) at Oak Ridge National Laboratory
(ORNL)~\cite{Efr05}, or in the LVD (Large Volume Detector)
experiment~\cite{Aga07}.

On the other hand, the cross sections $\nu_e(\bar{\nu}_e)-^{56}$Fe
are important to test the ability of
nuclear models in explaining reactions on nuclei with
masses around iron, which play an important role
in supernova collapse~\cite{Woo90}. The iron is
used as material detector in experiments on neutrino oscillations
such as MINOS~\cite{Ada07}, whereas future experiments,
such as SNS at ORNL \cite{Efr05} plan to use the same material.

There have been great efforts on nuclear structure models to
describe consistently  semileptonic weak processes with $^{12}$C
such as RPA-like models. A brief summary on the different models
employed for $^{12}$C is sketched in Ref.~\cite{Sam08a}.

The  puzzle with the Random Phase Approximation
(RPA) and the quasiparticle RPA (QRPA), when applied to the weak
observables in the triad
$\{{{^{12}{\rm B}},{^{12}{\rm C}},{^{12}{\rm N}}}\}$, is well known.
That is, to get agreement with data
for the ground state triplet $T=1$ ($\beta^\pm$-decays,
$\mu$-capture, and the exclusive $^{12}$C$(\nu_e,e^-)^{12}$N
reaction) the continuum RPA (CRPA) calculations
of Kolbe, Langanke, and Krewald~\cite{Kol94} needed
to be rescaled  by a reduction factor $\cong 4$.
The reason  for such a large discrepancy is very simple: within
the RPA the transitions $^{12}$C$\go^{12}$N$(1_1^+)$ and
$^{12}$C$\go^{12}$B$(1_1^+)$ are engendered mostly by the
particle-hole excitation  $p_{3/2}\go p_{1/2}$, what is physically
incorrect. In fact, since  late 1980's we know from  several hadronic
charge-exchange reaction measurements, and the consecutive Shell
Model (SM) calculations, that the
excitations $p_{3/2}\go p_{3/2}$,
$p_{1/2}\go p_{1/2}$, and  $p_{1/2}\go p_{3/2}$
participate  quite significantly in these  processes
(see, for instance,~\cite[Table I]{Win86}).
It is the involvement of these configurations that brings about
the necessary quenching of the Gamow-Teller (GT) resonances and
$\b$-decay rates. To make them come into play  it is mandatory
to open the $p_{3/2}$ shell by means of  pairing correlations,
which is done within both the SM and the QRPA.
But,  a new problem  emerges
in the  application of the QRPA to ${^{12}}$C, as first observed
by Volpe \etal\cite{Vol00} who noted that within this approach
the lowest state in ${^{12}}$N irremediable turned out not to be
the most collective one. As a consequence the QRPA also fails
in accounting for the exclusive processes
to the isospin triplet $T=1$. Soon after it was shown
\cite{Krm02,Krm05,Sam06} that the origin of this difficulty
arises from the degeneracy among   the ${p_{1/2}}$ and
${p_{3/2}}$ quasiparticle energies (both for protons and neutrons),
which is inherent to the non-conservation of particle number.
Therefore, for a physically sound description
of  the weak processes among  the $A=12$ iso-triplet it is
imperative to use the SM or the number projected QRPA (PQRPA).

The QRAP code is based on
Refs.~\cite{Krm02,Krm05,Sam06}, where a new  formalism
for neutrino-nucleus scattering has been developed,
and the PQRPA is used as the nuclear model framework.
The residual interaction was done with the
simple $\delta$-force, which  has been used extensively
in the literature to describe the single and double
beta decays~\cite{Hir90,Hir90a,Hir90b,Krm92,Krm93,Krm94}.

Before proceeding we   address briefly on the genesis of
the QRPA and PQRPA in  a manner
appropriate in the present context. Although this is not
a topic of central interest for the application-oriented
computer code, it belongs to the physics background.
The neutron-proton  QRPA was developed  in 1967 by
Hableib and Sorenson~\cite{Hab67} in order to account for
the hindrance  of the allowed  $\b$-transitions.
Almost 20 years later, when Vogel and Zirnbauer~\cite{Vog86}
and Cha~\cite{Cha87} discovered  the importance of the
particle-particle force in the S = 1, T = 0 channel, the
QRPA  became to be the most frequently used nuclear structure
method for evaluating double beta ($\b\b$) rates.
It was quickly realized, however, that a small change in
the particle-particle interaction  strength
caused a large change in the lifetimes and eventually
the breakdown (called a "collapse") of the entire method.
Later on several modifications of the QRPA were proposed to
make it more reliable. One of these was the charge-exchange
PQRPA, which has been formulated to evade the disadvantages
inherent in the non-conservation of particle number, and
was derived from the time-dependent variational
principle~\cite{Krm93}. But, the PQRPA   did not yield
substantially different result from the plain QRPA, and was
unable to avoid the collapse in  the study the two-neutrino
$\beta\beta$-decay in $^{76}$Ge. As a matter of fact, the
problem of the QRPA collapse has not yet been
settled down, in spite of enormous effort invested for
this purpose by many nuclear physicists (compare, for
instance,  Fig. 1 from Ref.~\cite{Krm93} with
Fig. 5 from a recent work of Yousef~\etal~\cite{You09}).

However, the PQRPA  turned out to be quite important
for the description of relatively light nuclei such
as $^{12}$C. For example, the employment of PQRPA for
the inclusive $^{12}$C$(\nu_e,e^-)^{12}$N cross section,
instead of the continuum RPA (CRPA) used by the LSND
collaboration in the analysis of ${\nu}_\mu \go{\nu}_e$
oscillations of the 1993-1995 data sample, leads to an
increased oscillation probability~\cite{Sam06}.

The PQRPA was recently also used to calculate the
$^{56}$Fe$(\nu_e,e^-)^{56}$Co cross section~\cite{Sam08}.
A comparison between the  QRPA and PQRPA for  the same interaction
and employing  the same model space  shows  that the projection
procedure could be important for medium mass nuclei. Moreover,
several approximations such as: i) Hybrid Model (HM)~\cite{Kol99},
ii)  QRPA with Skyrme interaction~\cite{Laz07}, iii) relativistic
QRPA (RQRPA)~\cite{Paa08}, and iv) QRPA and PQRPA with the
$\delta$-force~\cite{Sam08} yield different results for the
neutrino cross section as a function of the neutrino energy.
It is a hard task to find the origin for the differences,
mainly because these models are not using the same interaction
and/or the same single-particle configuration space, carrying
different types of correlations in each case.

The cross sections for charged- and neutral-current neutrino-induced
reactions on the iron isotopes $^{52-60}$Fe were also  evaluated
within the HM for  various supernova neutrino spectra~\cite{Toi01}.
Here, large-scale SM  calculations were used for the GT-like
contributions, while  transitions for other
multipoles are based on the RPA. More precisely, the authors scale
the SM cross sections using the ratios obtained from the  RPA
calculations with and without  this dependence of the multipole
operator. The  reason for such a procedure is twofold: i) the
limitation of the SM  to account for
momentum-transfer dependence of the GT operator, and ii) the lack
of pairing correlations in the RPA. It should be also mentioned
that SM  calculations of   inelastic neutral-current neutrino-nucleus
cross sections  in medium-mass nuclei, present in supernova
environment,   have been  constrained by the highly precise data on
the magnetic dipole strength distributions for the nuclei
$^{50}$Ti, $^{52}$Cr, and $^{54}$Fe, which
are dominated by spin-isospin flipping (GT-like)
contributions \cite{Lan04}. In spite of the agreement between data
and calculations it  was necessary to consider also here the effects
of finite momentum transfer what was  done via the RPA. Briefly,
the HM is neither fish nor fowl, and  a  comparison of the results
from   Refs.~\cite{Toi01,Lan04} with self-consistent  calculations,
such as the QRPA,  PQRPA and RQRPA, could be enlightening.

This brief introduction shows: 1) the importance of
neutrino-nucleus cross sections for astrophysical purposes and, 2)
that these cross sections are strongly correlated with the nuclear
structure model employed.  The QRAP code, with a simple residual
interaction, is able to access the sources of these problems and it
can calculate several weak interaction processes mentioned above.
Needless to stress that this code can be easily adapted for the
evaluation of $\b\b$-decays.

The write-up is organized as follows. In section II we make a short
survey of the theoretical description of weak interaction processes,
with emphasis on the formulation implemented in this numerical code.
In sections III and IV we describe the QRPA, and PQRPA
formalisms, making explicit the differences among them. In section V
we show how the code is organized, how to make an input and how to
understand the output. Section VI explains the role of each
subroutine of the code. Finally, section VII proposes a few cases to
practice with the code.

\section{Weak interacting processes}

In this section we give a brief summary of the main  formulae
developed in Ref.~\cite{Sam08a,Krm05} for:
\bit \item neutrino scattering (NS)
\brn \nu_\ell + (Z, A)&\go& (Z+1,N-1)+\ell^-,
\ern
\item  antineutrino scattering (AS)
\brn {\bar \nu}_\ell + (Z,A) &\go& (Z-1, N+1)+\ell^+,
\ern
\item muon capture (MC) rate
\brn
\mu^- + (Z, A)&\go& (Z-1, N-1)+ \nu_\mu,
\ern
\eit where $\ell= e, \mu$.
The comparison with other formalisms~\cite{Don79,Wal04,Kur90}
can be found is in just mention works.

The weak Hamiltonian is expressed in the form %\cite{Don79,Wal04,Bli66}
\br
H_{{\sss {W}}}(\rb)&=&\frac{G}{\sq}J_\alpha l_\alpha
e^{-i\rb\cdot\kb},
\label{2.1}\er
where $G=(3.04545\pm 0.00006){\times} 10^{-12}$ is
the Fermi coupling constant (in natural units),
\br
J_\alpha&\equiv& (\Jb,iJ_\emptyset)
\label{2.2}\\
&=&i\g_4
\left[\gV\g_\alpha-\frac{\gM}{2\Mass}\s_{\alpha\beta}k_\beta
+\gA\g_\alpha\g_5+i\frac{\gP}{\mass_\ell} k_\alpha\g_5\right],
\nn\er
 is the hadronic current
operator\footnote{To avoid confusion, we will be using roman
fonts ($\Mass$,$\mass$) for masses and math italic fonts ($M$,$m$)
for azimuthal quantum numbers.},  and \br
l_\alpha(\qb,E_\nu)\equiv (\lb,il_\emptyset)
&=&-i\overline{u}_{s_\ell}(\pb,E_\ell)\g_\alpha(1+\g_5)u_{s_\nu},
\nn\\
\label{2.3}\er
is the plane wave
approximation for the matrix element of the leptonic current in
the case of neutrino reactions, with $p_\ell\equiv\{\pb,
iE_\ell\}$ and $q_\nu\equiv\{\qb,iE_{\nu}\}$ being, respectively,
the lepton and the neutrino momenta.

For the sake of convenience we will use spherical coordinates
($m=-1,0,+1$) for the three-vectors, and the Walecka's notation
\cite{Wal04}, with the Euclidean metric, for four-vectors,
 \ie  $x= \{ \xb,x_4=i x_\emptyset \}$. The only difference
is that we substitute Walecka's indices $(0,3)$ by our indices
$(\emptyset,0)$, i.e. we use the index $\emptyset$ for the
temporal component and the index $0$ for the third spherical
component.

The quantity
\br k = P_i-P_f\equiv \{\kb,ik_\emptyset \},
\label{2.4}\er
 is the momentum transfer, where $P_i$ and $P_f$ are
momenta of the initial and final nucleus, ${\rm M}$ is the nucleon
 mass, ${\rm m}_\ell$ is the mass of the charged
lepton, and $g_{\sss V}$, $g_{\sss A}$, $g_{\sss M}$ and $g_{\sss
P}$ are, respectively, the vector, axial-vector, weak-magnetism
and pseudoscalar effective dimensionless coupling constants. Their
numerical values are:
\br
 g_{\sss V}&=&1 ;
~g_{\sss A}=1.26 ; \nn\\
g_{\sss M}&=&\kappa_p-\kappa_n=3.70 ; ~g_{\sss P}= g_{\sss
A}\frac{2\Mass \mass_\ell }{k^{2}+\mass_\pi^2}.
\label{2.5} \er
In the numerical calculations we use an effective axial-vector
coupling $g_{\sss A}^{\sss}=1$ \cite{Cas87}.

The finite nuclear size (FNS) effect is incorporated via the
dipole form factor with a cutoff $\Lambda=850$ MeV, \ie
\be g\go
g\left( \frac{\Lambda^{2}}{\Lambda^{2}+k^{2}}\right)^{2}.
\label{2.6} \ee

To use \rf{2.1} with  the non-relativistic nuclear wave functions,
the Foldy-Wouthuysen transformation has to be performed on the
hadronic current \rf{2.2}. When the velocity dependent terms are
included this yields \cite{Bli66}:
\br
 J_\emptyset&=&\gV + (\ga+\gpa) {\mbs} \cdot\hat{\kb} - \gA
\mbs \cdot \vb,
\nonumber\\
\Jb&=& -\gA {\mbs} -i\gw {\mbs}\x\hat{\kb}-\gv \hat{\kb}
+\gpb({\mbs} \cdot\hat{\kb})\hat{\kb}+{\gV}\vb,
\nn\\
\label{2.7}\er
where $\hat{\kb}=\kb/\k$, $\k\equiv\absk$, and
 $\vb\equiv -i \mbn/\Mass$ is the velocity operator, acting on the
nuclear wave functions. The following short notation
\br
\gv&=&\gV\frac{\k}{2\Mass};~
\ga=\gA\frac{\k}{2\Mass};~ \gw=(\gV+\gM)\frac{\k}{2\Mass},
\nn\\
\gpa&=&\gP\frac{\k}{2\Mass}\frac{q_\emptyset}{\mass_\ell};~
\gpb=\gP\frac{\k}{2\Mass}\frac{\k}{\mass_\ell},
\label{2.8}\er
has also been introduced.

In performing the multipole expansion of the nuclear operators
\be O_\alpha\equiv
(\Ob,O_\emptyset)=J_\alpha e^{-i\kr},
\label{2.9}\ee
it is convenient:

1) to take the momentum $\kb$ to be along  the $z$ axis,
\ie
\br
e^{-i\kr}&=&\sum_{\sf  L}i^{-\sf  L}\sqrt{4\pi(2{\sf L}+1)}
j_{\sf  L}(\rho) Y_{{\sf L}0}(\hat{\rb}),
\nn\\
&=&\sum_{\sf  J}i^{-\sf J}\sqrt{4\pi(2{\sf J}+1)}
j_{\sf  J}(\rho) Y_{{\sf J}0}(\hat{\rb}),
\label{2.10}\er
where $\rho=\k r$, and

2) to introduce  the operators ${\sf O}_{\a{\sf J}}$, defined as
\be
O_\alpha\equiv ({\bf O}, O_{\emptyset})=\sqrt{4\pi}
\sum_{\sf  J}i^{-{\sf J}}\sqrt{2{\sf J}+1}{\sf O}_{\a{\sf J}}.
\label{2.11}\ee
Thus,
\br
{\sf O}_{\emptyset{\sf J}}&=&
j_{\sf J}(\rho)Y_{{\sf J}0}(\hat{\rb}) J_\emptyset,
\nn\\
{\sf O}_{m{\sf J}}&=&\sum_{{\sf L}}i^{\sf  J-L} F_{m\sf LJ}j_{\sf
L}(\rho) \left[Y_{{\sf L}}(\hat{\rb})\otimes{\Jb}\right]_{\sf J },
\nn\\
\label{2.12}\er
 where the geometrical factors
\br F_{m\sf JL}&\equiv& (-) ^{m+\sf J}\sqrt{(2{\sf L}+1)}
\threej{{\sf L}}{1}{{\sf J}}{0}{-m}m, \label{2.13}\er are listed
in Table I of Ref.~\cite{Krm05}.

Explicitly, from \rf{2.7}
\br
{O}_{\emptyset{\sf J}}&=&g_{\sss{V}}\M_{\sf J}^{\sss V}+i\gas\M^{\sss A}_{\sf J}
+i(\ga+\gpa)\M^{\sss A}_{0{\sf J}}
\label{2.14}\\
{O}_{{m}{\sf J}} &=&i(\delta_{{m}0}\gpb-\gA +m \gw)\M^{\sss A}_{{m}{\sf J}}
\nn\\
&+&\gvs\M^{\sss V}_{{m}{\sf J}}-\delta_{{m} 0}\gv\M_{\sf J}^{\sss V}.
\label{2.15}\er

The elementary operators are given by
 \br
  \M^{\sss V}_{\sf J}&=&j_{\sf J}(\rho) Y_{{\sf J}}(\hat{\rb}),
\nn\\
\M^{\sss A}_{\sf J}&=&
{\rm M}^{-1}j_{\sf J}(\rho)Y_{\sf J}(\hat{\rb})(\mbs\cdot\mbn),
\nn\\
\M^{\sss A}_{{m\sf J}}&=&\sum_{{\sf L}\ge 0}i^{ {\sf J-L}-1}
\F_{{m\sf L\sf J}}j_{\sf L}(\rho)
\left[Y_{{\sf L}}(\hat{\rb})\otimes{\mbs}\right]_{{\sf J}},
\label{2.16}\\
\M^{\sss V}_{{m\sf J}}&=&{\rm M}^{-1}\sum_{{\sf L}\ge 0}i^{ {\sf J-L}-1}
F_{{m\sf L\sf J}}j_{\sf L}(\rho) [ Y_{\sf L}(\hat{\rb})\otimes\mbn]_{{\sf J}}.
\nn\er

Here we make use of the conserved vector current (CVC). From
\rf{2.14}, \rf{2.15}, and \cite[Eq. (10.45) and (9.7)]{Beh82}
\br
\kb\cdot\Ob^{\sss V}=\k O_0^{\sss V}= k_{\emptyset}
O_\emptyset^{\sss V} \label{2.17}\er which yields \br
g_{\sss{V}}\M^{\sss V}_{{\sf 0}{\sf J}}-\gv\M_{\sf J}^{\sss V}
=\frac{k_{\emptyset}}{\k}g_{\sss{V}} \M^{\sss V}_{\sf J}.
\label{2.18}\er
Therefore, from \rf{2.15} \br {\sf O}_{m{\sf J}}
&=&i(\delta_{m0}\gpb-\gA +m\gw)\M^{\sss A}_{m{\sf J}}
\nn\\
&+&2|m|\gv\M^{\sss V}_{m{\sf J}} +\delta_{m
0}\frac{k_{\emptyset}}{\k}\gV \M_{\sf J}^{\sss V}. \label{2.19}\er

The elementary operators
$\M^{\sss V}_{\sf J}$, $\M^{\sss A}_{\sf J}$, $\M^{\sss A}_{\sf 0J}$
and $\M^{\sss V}_{\sf 0J}$ are real,
but $\M^{\sss A}_{\pm1{\sf J}}$ and $\M^{\sss V}_{\pm1{\sf J}}$
are not, and it is  convenient to put in evidence their real and
imaginary parts, expressing  them as
\br
\M_{\pm1{\sf J}}&=&{\M}^{\sss R}_{1{\sf J}}\pm i{\M}^{\sss I}_{1{\sf J}}
\label{2.20}\er
with ${\M}^{\sss  R}_{1{\sf J}}$,
and  ${\M}^{\sss  I}_{1{\sf J}}$ arising, respectively,  from the terms
in \rf{2.16} with ${\sf L}= {\sf J}\pm1 $, and ${\sf L}= {\sf J}$.
Note that  $F_{\pm1\sf JJ}=\mp 1/\sqrt{2}.$

It is also convenient to separate the elementary operators into:
\bit
\item {\it natural parity} (NP),
($\pi=(-)^{\sf J}$):
$\M^{\sss V}_{\sf J}$, ${\M}^{\sss  A,I}_{1{\sf J}}$,
and ${\M}^{\sss V,R}_{1{\sf J}}$, and
\item {\it unnatural parity} (UP), ($\pi=(-)^{{\sf J}+1}$):
$\M^{\sss A}_{\sf J}$,  ${\M}^{\sss  V,I}_{1{\sf J}}$,
$\M^{\sss A}_{{0 \sf  J}}$,
and ${\M}^{\sss A,R}_{1{\sf J}}$
\eit

The operators ${\sf O}_{\a{\sf J}}\equiv ({\sf O}_{\emptyset{\sf
J}}, {\sf O}_{m{\sf J}})$ can be express as a sum of real and
imaginary operators, \ie ${\sf O}_{\a{\sf J}}={\sf O}^{\sss R}_{\a{\sf
J}}+i{\sf O}^{\sss I}_{\a{\sf J}}$, with  ${\sf O}^{\sss R}_{\a{\sf J}}$
(${\sf O}^{\sss I}_{\a{\sf J}}$) being a NP (UP) operator.  This is a
very important finding because it implies that ${\sf
O}^{\sss R}_{\a{\sf J}}$ and ${\sf O}^{\sss I}_{\a{\sf J}}$
{\em  do not contribute simultaneously}, and,  therefore,
{\em one always can deal only with real operators.}

In summary, natural and unnatural parity operators
are, respectively:
\br
{\sf O}^{\sss R}_{\emptyset{\sf J}}&=&g_{{\sss{V}}}\M_{\sf J}^{\sss V},
\nn\\
{\sf O}^{\sss R}_{{0}{\sf J}} &=&\frac{k_{\emptyset}}{\k}
\gV\M_{\sf J}^{\sss V},
\nn\\
{\sf O}^{\sss R}_{m\ne 0{\sf J}}
&=&(m\gA -\gw) {\M}^{\sss A,I}_{1{\sf J}}+\gvs{\M}^{\sss V,R}_{1{\sf J}},
\label{2.21}\er
and \br {\sf O}^{\sss I}_{\emptyset{\sf J}}&=&\gas\M^{\sss A}_{\sf J}
+(\ga+\gpa)\M^{\sss A}_{0{\sf J}},
\nn\\
{\sf O}^{\sss I}_{{0}{\sf J}} &=&(\gpb-\gA )
\M^{\sss A}_{{0 \sf  J}},
\nn\\
{\sf O}^{\sss I}_{m\ne 0{\sf J}} &=&(-\gA +m\gw){\M}^{\sss A,R}_{1{\sf J}}
+\gvs{\M}^{\sss V,I}_{1{\sf J}}. \label{2.22}\er

\subsection{Neutrino-nucleus cross section}

For the neutrino-nucleus reaction, the momentum transfer is
$k=p_\ell-q_\nu$, and the corresponding cross section reads \br
\s(E_\ell,J_f)& = &\frac{|\pb_\ell| E_\ell}{2\pi} F(Z\pm1,E_\ell)
\int_{-1}^1
d(\cos\theta)\T_{\s}(\q,J_f),\nn\\
\label{2.23}\er
where $F(Z\pm1,E_\ell)$ is  the Fermi function
($Z+1$, for neutrino, and $Z-1$, for antineutrino),
$\theta\equiv\hat{\qb}\cdot\hat{\pb}$ is the angle between the incident
neutrino and ejected lepton, and the transition amplitude is
\br
\T_{\s}(\k,J_f)= \frac{1}{2J_i+1} \sum_{ s_\ell,s_\nu }\sum_{M_i, M_f }
\left|\bra{J_fM_f}H_{{\sss {W}}}\ket{J_iM_i}\right|^{2}.
\nn\\
\label{2.24}\er
After   expressing  the spatial part of the lepton
traces $\L_{\a\b}$ in spherical coordinates, and   applying  the
Wigner-Eckart theorem, one can cast the transition amplitude in
the compact form~\cite{Krm05}
\br
\T_{\s}(\k,J_f)&=&\frac{4\pi
G^2}{2J_i+1}\sum_{{\sf J}}\left[ |\Bra{J_f}{\sf O}_{\emptyset{\sf
J}}\Ket{J_i}|^2\L_{\emptyset} \right.
\nn\\
&+&\sum_{m=0\pm 1}|\Bra{J_f}{\sf O}_{m{\sf  J}}\Ket{J_i}|^2\L_m
\label{2.25}\\
&-&\left.2\Re(|\Bra{J_f}{\sf O}_{\emptyset{\sf J}}\Ket{J_i}
\Bra{J_f}{\sf O}_{0{\sf J}}\Ket{J_i})\L_{\emptyset 0}\right].
\nn\er The explicit expressions for the traces
$\L_{\emptyset}\equiv \L_{\emptyset\emptyset}$, $\L_m\equiv
\L_{mm}$, and $\L_{\emptyset 0}$ are~\cite{Krm05} \br
\L_{\emptyset\emptyset}&=&1+\frac{|\pb|\cos\theta}{E_\ell},
\nn\\
\L_{\emptyset
0}&=&\left(\frac{q_0}{E_\nu}+\frac{p_0}{E_\ell}\right),
\nn\\
\L_{0}&=&1+\frac{2q_0p_0}{E_\ell
E_\nu}-\frac{|\pb|\cos\theta}{E_\ell},
\nn\\
\L_{\pm1}&=&1-\frac{q_0p_0}{E_\ell E_\nu}\pm
\left(\frac{q_0}{E_\nu}-\frac{p_0}{E_\ell}\right) S_1,
\label{2.26}\er
 with \br q_0&=&{\hat k}\cdot
\qb=\frac{E_\nu(|\pb|\cos\theta-E_\nu)}{\k},
\nn\\
p_0&=&{\hat k}\cdot \pb=\frac{|\pb|(|\pb|-E_\nu\cos\theta)}{\k},
\label{2.27} \er
being the $z$-components of the neutrino and lepton
momenta, and $S_1=\pm 1$ for NS and AS, respectively.

\subsection{$\mu$-capture rates}

The muon capture transition amplitude $\T_{\sss MC}(J_f)$ can be
derived from the result \rf{2.25} for the neutrino-nucleus reaction
amplitude, by keeping  in mind that: i) the roles of $p$ and $q$
are interchanged within  the
matrix elements of the leptonic current, which makes that
in \rf{2.26} $S_1\go -1$, ii) the momentum
transfer turns out to be $k=q-p$, and therefore the signs on the
right-hand sides of ($q_0, p_0$) have to be changed, and iii) the
threshold values ($\pb\go 0: \qb \go \kb, k_\emptyset\go
E_\nu-\mass_\ell$) must be used for the lepton traces. All this
yields $q_0=E_\nu$, $p_0=0$, and
\be
\L_{\emptyset\emptyset}=\L_{\emptyset0}=\L_{0}=1,~~\L_{1}=
0,~~\L_{-1}=2.
\label{2.28}\ee
Instead of summing over the initial
lepton spins $s_\ell$, as done in \rf{2.24}, one has now to average
over the same quantum number. We get
 \br
\Lambda(J_f)&=&\frac{E_\nu^2}{2\pi}|\phi_{1S}|^2\T_{\sss MC}(J_f),
\label{2.29}\er
 where $\phi_{1S}$ is the muonic bound state wave
function evaluated at the origin, and
$E_\nu=\mass_\mu-(\Mass_n-\Mass_p)-E_B^\mu-E_f+E_i$, where
$E_B^\mu$ is the binding energy of the muon in the $1S$ orbit.
Thus from \rf{2.25} and \rf{2.28}
\br
\T_{\s}(\k,J_f)&=&\frac{4\pi
G^2}{2J_i+1}\sum_{{\sf J}}\left[ |\Bra{J_f}{\sf O}_{\emptyset{\sf J}}
-{\sf O}_{0{\sf J}}\Ket{J_i}|^2 \right.
\nn\\
&+&\left. 2|\Bra{J_f}{\sf O}_{-1{\sf J}}\Ket{J_i}|^2\right].
\label{2.30}\er

In the case of MC it
is convenient to rewrite the effective coupling constants \rf{2.8}
as
\br \gv&=&\gV\frac{E_\nu}{2\Mass};~
\ga=\gA\frac{E_\nu}{2\Mass};~
\nn\\
\gw&=&(\gV+\gM)\frac{E_\nu}{2\Mass};~~
\gp=\gP\frac{E_\nu}{2\Mass},
\label{2.31}\er
where $\gp=\gpb-\gpa$.
\footnote{Note that there is a misprint in  Eq. (2.41) of
Ref.~\cite{Krm05}. Also in  Eq. (2.42) of the same reference
$\gpa$ should read $\gp$.}

Thus, natural and
 unnatural parity operators are now, respectively:
\br
{\sf O}^{\sss R}_{\emptyset{\sf J}}-{\sf O}^{\sss R}_{0 \sf  J}&=&
(\gV-\frac{k_{\emptyset}}{\k}\gV )\M^{\sss V}_{\sf  J}
=\gV\frac{\mass_\mu}{E_\nu}\M^{\sss V}_{\sf  J},
\nn\\
{\sf O}^{\sss R}_{-1{\sf J}} &=&-(\gA +\gw)
{\M}^{\sss A,I}_{1{\sf J}}+\gvs{\M}^{\sss V,R}_{1{\sf J}},
\label{2.32}\er
and
\br
{\sf O}^{\sss I}_{\emptyset{\sf J}}-{\sf O}^{\sss I}_{0 \sf  J}&=&
\gas\M^{\sss A}_{\sf J}+
\left(\gA  +\ga-\gp\right)\M^{\sss A}_{0{\sf J}},
\nn\\
\sf O}^{\sss I}_{-1{\sf J} &=&-(\gA +\gw){\M}^{\sss A,R}_{1{\sf J}}
-\gvs{\M}^{\sss  V,I}_{1{\sf J}}.
\label{2.33}\er

\section{Nuclear structure calculation}

\subsection{PQRPA}

The  PQRPA for
charge-exchange excitations was derived from the time-dependent
variational principle  in Ref.~\cite{Krm93}. In the same reference
is also  described in details the projected  Barden-Cooper-Schiffer
(PBCS) approximation.
Basically one  employs
the number projection operators ${\hat P}_\N$ on the $\ket{BCS}$
state. That is: ${\hat P}_0={\hat P}_Z{\hat P}_N$
for a ground state with $Z$ protons and $N$ neutrons,
and ${\hat P}_\mu={\hat P}_{Z+\mu}{\hat P}_{N-\mu}$,
with  $\mu=\pm 1$, for  excited states in nuclei with $Z+\mu$ protons
and $N-\mu$ neutrons. In this section we give
a brief description of both the PBCS and PQRPA approximations.

The PBCS gap equations are
\br
2\bar{e}_k u_k v_k-\Delta_k(u^2_k-v^2_k)&=&0,
\label{3.1}\er
where
\br
&&\Delta_k=-\frac{1}{2}\sum_{k'}{(2j_{k'}+1)^{1/2}\over
(2j_k+1)^{1/2}}u_{k'}v_{k'}
{\rm G}(kkk'k';0)\frac{I^{Z-2}(kk')}{I^Z},
\nn\\
\label{3.2}\er
are the pairing gaps, and
\br
{\bar e}_k &=& e_k \frac{I^{Z-2}(k)}{I^Z} +
\sum_{k'} {(2j_{k'}+1)^{1/2}\over (2j_k+1)^{1/2}} v_{k'}^2
\nn\\
&&\times {\rm F}(kkk'k';0)\frac{I^{Z-4}(kk')} {I^Z}+ \Delta
e_k,\label{3.3}\er
are  the dressed  single-particle energies,
where
\br
I^{K}(k_1k_2\cdot\cdot k_n)&=& \frac{1}{2\pi i}
\oint \frac{dz}{z^{K+1}} \s_{k_1}\cdots \s_{k_n}
\nn\\
 &\x&\prod_{k}(u_k^2 + z^2 v_k^2)^{j_k+1/2};
 \nn\\
 \s_{k}^{-1}&=&u^2_k+z^2_kv_k^2,
\label{3.4}\end{eqnarray}
are the  PBCS number projection integrals.
The PBCS correction term $\Delta e_k$ can be
found in Ref.~\cite{Krm93},
 ${\rm G}(kkk'k';0)$, and ${\rm F}(kkk'k';0)$ stand for the
usual proton or neutron  particle-particle (pp), and
particle-hole (ph)  matrix  elements of the residual
interaction $V$, \ie
\br
G(klmn;J)&=&\bra{kl;J}V\ket{mn;J}.
\nn\\
F(klmn;J)&=&\bra{kl^{-1};J}V\ket{mn^{-1};J}.
\label{3.5}\er
Note that these relations are valid for both identical and
non-identical particles.

The forward-going ($X_{\mu}$), and backward-going
($Y_{\mu}$)  PQRPA amplitudes are obtained by solving
the RPA equations
\begin{eqnarray}
\left(\begin{array}{cc} A_\mu &  B \\  -B^{*}&
 -A^{*}_{-\mu} \end{array}\right) \left(\begin{array}{l}
X_\mu \\ Y_\mu  \end{array}\right) = \omega_\mu
 \left(\begin{array}{l} X_\mu  \\
Y_\mu  \end{array}\right),
\label{3.6}\end{eqnarray}
with the PQRPA matrices defined as:
\br
&&A_\mu(pn,p'n';J)
\nn\\
&&=
\omega^0_\mu\del_{pn,p'n'}
+ N^{-1/2}_\mu(pn)N^{-1/2}_\mu(p'n')
\nn\\
&&
\times\left\{[u_pv_nu_{p'}v_{n'}I^{Z-1+\mu}(pp')I^{N-3-\mu}(nn')
\right.
\nn\\
&&+
v_pu_nv_{p'}u_{n'}I^{Z-3+\mu}(pp')I^{N-1-\mu}(nn')]
{\rm F}(pn,p'n';J)\nn\\
&&+
[u_pu_nu_{p'}u_{n'}I^{Z-1+\mu}(pp')I^{N-1-\mu}(nn')
\nn\\
&&+
\left.v_pv_nv_{p'}v_{n'}I^{Z-3+\mu}(pp')I^{N-3-\mu}(nn')]
{\rm G}(pn,p'n';J)\right\}, \nn\\
&&B(pn,p'n';J)
\nn\\
&&=N^{-1/2}_\mu(pn) N^{-1/2}_{-\mu}(p'n')I^{Z-2}(pp')I^{N-2}(nn')
\nonumber\\
&&\times [(v_pu_nu_{p'}v_{n'}+u_pv_nv_{p'}u_{n'})
{\rm F}(pn,p'n';J)
\nn\\
&&+(u_pu_nv_{p'}v_{n'}+v_pv_nu_{p'}u_{n'})
{\rm G}(pn,p'n';J)],
\label{3.7}\er
where
\be
\omega^0_\mu=\varepsilon^{Z-1+\mu}_p+\varepsilon^{N-1-\mu}_n
\label{3.8}\ee
are the unperturbed energies,
\br
N_\mu(pn)=I^{Z-1+\mu}(p)I^{N-1-\mu}(n),
\label{3.9}\er
are the norms,
\br
\varepsilon^{K}_k=
{R_0^K(k)+R_{11}^K(kk)\over I^K(k)}-\frac{R_0^K}{I^K}
\label{3.10}\er
are the projected quasiparticle energies, and
the quantities $R^K$ are defined as \cite{Krm93}
\br
R_0^K(k)&=&\sum_{k_1} (2j_{k_1}+1)v_{k_1}^2e_{k_1}I^{K-2}(k{k_1})
\nn\\
&+&\frac{1}{4}\sum_{{k_1}{k_2}}(2j_{k_1}+1)^{1/2}(2j_{{k_2}}+1)^{1/2}
\nn\\
&\x&\left[v_{k_1}^2v_{{k_2}}^2{\rm F}({k_1}{k_1}{k_2}{k_2};0)
I^{K-4}({k_1}{k_2}k)\right.
\nn\\
&+&\left.u_{k_1}v_{{k_1}}u_{{k_2}}v_{{k_2}}
{\rm G}({k_1}{k_1}{k_2}{k_2};0) I^{K-2}({k_1}{k_2}k)\right],
\nn\\
R_{11}^K(kk)&=&e_{k}[u_{k}^2 I^{{K}}(kk)-v_{k}^2 I^{K-2}(kk)]
\nn\\
&+&\sum_{{k_1}}{(2j_{k_1}+1)^{1/2}\over(2j_{k}+1)^{1/2}}
\left\{v_{k_1}^2{\rm F}({k_1}{k_1}kk;0) \right.
\nn\\
&&[u_{k}^2I^{K-2}({k_1}kk)-v_{k}^2I^{K-4}({k_1}kk)]
\nn\\
&&-\left.u_{k_1}v_{{k_1}}u_{k}v_{k}{\rm G}({k_1}{k_1}kk;0)
I^{K-2}({k_1}kk)\right\}.
\nn\\
\label{3.11}\er

Both positive and the negative solutions are physically
meaningful. For $\mu= \pm1$ the positive solutions describe
excitations in the $(Z\pm1, N\mp1)$ nuclei,
while the negative energy solutions represent the positive
energy excitations in the $(Z\mp1, N\pm1)$. Thus only one RPA
equation has to be solved, either for $\mu=+1$, or for $\mu=-1$,
to describe the  excitations to the $Z\pm1,N\mp1$ nuclei.
This is   well known feature  of the charge-exchange
modes~\cite{Row70,Lan80,Boh75,Krm80}.

Let us be  more specific, and take advantage of the index $f$
to label different final states $\ket{J_f^\pi}$ with same spin and parity.
Evidently, $f$ will run from $1$ up the total number  $f_{\rm max}$
of two-quasiparticle configurations $\ket{pnJ^\pi}$. Moreover,
the eigenvalue problem \rf{3.6} has  $2f_{\rm max}$ solutions,
and  we will use the index $F$ to label them.
Thus, if $\mu=+1$ one have:

$\bullet$ {for $\w_{+1}(J_{F})>0$~($1<F\le f_{\rm max}$):
\br
\w_{+1}(J_f)&=&\w_{+1}(J_{F}),
\nn\\
X_{+1}(pnJ_{f})&=& X_{+1}(pnJ_{F}),
\nn\\
Y_{+1}(pnJ_{f})&=& Y_{+1}(pnJ_{F});
\label{3.13}
\er

$\bullet$ {for $\w_{+1}(J_{F})<0$~($f_{\rm max}<F\le 2f_{\rm max}$):
\br
\w_{-1}(J_f)&=&-\w_{+1}(J_{F}),
\nn\\
X_{-1}(pnJ_{f})&=& Y^*_{+1}(pnJ_{F}),
\nn\\
Y_{-1}(pnJ_{f})&=& X^*_{+1}(pnJ_{F}).
\label{3.14}
\er
Finally,  to store the eigenvalues and eigenfunctions it is
convenient to define  the index $F$ as
\begin{eqnarray}
F&=&
\left\{
\begin{array}{l}
f, \hspace{2.5cm}\mbox{for}\hspace{0.5cm} F\le f_{\rm max};
\\
2f_{\rm max}-f+1, \hspace{0.6cm}\mbox{for}\hspace{0.5cm} F> f_{\rm max}.
\\\end{array}\right.
\label{3.12}
\end{eqnarray}

\subsection{QRPA}

The usual gap equations are obtained from
Eqs. \rf{3.6}-\rf{3.7} by:
\begin{enumerate}
\item  Making  the replacement $e_{k}\rightarrow e_{k} -
\lambda_{\rm k}$, with $\lambda_{\rm k }$ being the chemical
potential, and taking the limit  $I^K \go 1$.
That is, the Eq. \rf{3.1} remains as it is, but instead
of \rf{3.2} and \rf{3.3} one has now
\begin{eqnarray}
&&\Delta_k=-\frac{1}{2}\sum_{k'}{(2j_{k'}+1)^{1/2}\over
(2j_k+1)^{1/2}}
u_{k'}v_{k'}
{\rm G}(kkk'k';0),
\nn\\\label{3.15}\er
and
\br
{\bar e}_k &=& e_k-\l_{\rm k}+
\sum_{k'} {(2j_{k'}+1)^{1/2}\over (2j_k+1)^{1/2}} v_{k'}^2 {\rm
F}(kkk'k';0).
\nn\\\label{3.16}\er
\item
Impose  the subsidiary conditions
\br
Z = \sum_{j_p} (2j_p+1)^2 v_{j_p}^2,
~~ N = \sum_{j_n} (2j_n+1)^2 v_{j_n}^2,
\label{3.17}\er
 as the number of particles is not any more
a good quantum number.
\end{enumerate}

In this way the usual BCS gap equations read
\be 2(e_k-\l_t)u_kv_k=(u_k^2-v_k^2)\Delta_k.
\label{3.18}\ee
This equation, together  with the
normalization condition $u_k^2+v_k^2=1$, has as solution the
occupation probabilities (for example, from the Chapter I of
Rowe~\cite{Row70})
\be
u_k^2=\frac{1}{2}\left(1+\frac{e_k-\l_k}{E_k}\right),~~~~~
v_k^2=\frac{1}{2}\left(1-\frac{e_k-\l_k}{E_k}\right),
\label{3.19}\ee
which depend on  the quasiparticle energies
\be
E_k=\sqrt{(e_k-\l_k)^2+\Delta_k^2},
\label{3.20}\ee and the
pairing gaps
\be
\Delta_k=-\frac{1}{2}\sum_{k'}{(2j_{k'}+1)^{1/2}\over
(2j_k+1)^{1/2}} u_{k'}v_{k'} {\rm G}(kkk'k';0).
\label{3.21}\ee

The QRPA equations are recovered from \rf{3.6}
by i) dropping the index $\mu$,  ii) taking
the limit $I^K \go 1$, and iii) substituting
the unperturbed  PBCS energies
by the BCS energies relative to the Fermi level,
defined by equation
\be
E^{(\pm)}_k=\pm E_{k}+\l_{\rm k},
\label{3.22}\ee
where ${E}_{k}$
are the usual BCS quasiparticle energies defined in \rf{3.20}.
In this way the unperturbed energies
in \rf{3.7} are replaced  by
\be
\omega^0_\mu= E_{j_p}+E_{j_n}+\mu(\l_{\rm p}-\l_{\rm n}).
\label{3.23}\ee
These energies, however, are not used in the QRPA eigenvalue
problem. Namely, the coefficients
$X(pnJ_f)$ and $Y(pnJ_f)$, and the eigenvalues
$\w(J_f)$ are obtained from
\be
\left(\begin{array}{cc}{\cal A}&{\cal B}\\{\cal B}&{\cal
A}\\\end{array}\right)
\left(\begin{array}{c}X\\Y\\\end{array}\right)=\w
\left(\begin{array}{c}X\\-Y\\\end{array}\right), \label{3.24}\ee
where
\br {\cal A}(pnp'n';J)
&=&(E_p+E_n)\delta_{pp'}\delta_{nn'}\nn\\
&+&(u_pv_nu_{p'}v_{n'}+v_pu_nv_{p'}u_{n'})F(pnp'n';J)\nn\\
&+&(u_pu_nu_{p'}u_{n'}+v_pv_nv_{p'}v_{n'})G(pnp'n';J),\nn\\
{\cal B}(pnp'n';J)
&=&(v_pu_nu_{p'}v_{n'}+u_pv_nv_{p'}u_{n'})F(pnp'n';J)
\nn\\
&+&(u_pu_nv_{p'}v_{n'}+v_pv_nu_{p'}u_{n'})G(pnp'n';J).
\nn\\\label{3.25}\er

In the pn-QRPA  the  eigenvalues occur in pairs
$\pm \w(J_f)$, but the negative energies don't have
a direct physical meaning.
The perturbed energies for  daughter
$(Z+\mu,N-\mu)$  nuclei  are defined as
\be
\omega_\mu(J_f)=\w(J_f)+\mu(\l_p-\l_n).
\label{3.26}\ee
There is, however,  only one set of eigenfunctions
$(X(pnJ_f), Y(pnJ_f))$ for both $\mu=1$, and $\mu=-1$.
This is a very important difference in relation to the
PQRPA case, which is crucial for the distribution
of the transition strengths.

\subsection{Nuclear Matrix Elements}

When the excited states $\ket{J_f}$ in the final $(Z\pm 1,N\mp 1)$
nuclei are described within the PQRPA, the transition amplitudes
for the multipole charge-exchange operators~\rf{2.21} and \rf{2.22} read
%${\sf Y}_{J}$, \etc, listed in Table II of Ref.~\cite{Krm05}, read
\br
&\Bra{J_f, Z+\mu,N-\mu}{\sf O}_{{\sf J}}\Ket{0^+}\nn\\
& = {1 \over (I^{Z}I^N)^{1/2}}
\sum_{pn}
 \left[ {\Lambda_\mu(pnJ) \over
 (I^{Z-1+\mu}(p)I^{N-1+\mu}(n))^{1/2}}
 X_{\mu}^{\ast}(pnJ_f)\right.\nonumber\\
& +\left.{\Lambda_{-\mu}(pnJ) \over
 (I^{Z-1-\mu}(p)I^{N-1-\mu}(n))^{1/2}} Y_\mu^{\ast}(pnJ_f)\right],
\label{3.27} \er
with the one-body matrix elements given by
\begin{eqnarray}
&&\Lambda_\mu(pnJ)=-\frac{\Bra{p}{\sf O}_{{\sf J}}\Ket{n}}{\sqrt{2J+1}}
\left\{
\begin{array}{l}
\up \vn,\;\; \mbox{for}\; \mu = +1 \\ \un \vp,\;\; \mbox{for}\; \mu
= -1 \
\\\end{array}\right.,
\nn\\
\label{3.28}
\end{eqnarray}
and $J={\sf J}$.

In the QRPA case, using the limit $I^K \go 1$ in \rf{3.27},
the nuclear matrix elements for the
multipole charge-exchange operators ${\sf O}_{{\sf J}}$
are
\br
&\Bra{J_f, Z+\mu,N-\mu}{\sf O}_{{\sf J}}\Ket{0^+}\nn\\
& =
\sum_{pn}
 \left[ \Lambda_\mu(pnJ)
 X^{\ast}(pnJ_f)
 +\Lambda_{-\mu}(pnJ)  Y^{\ast}(pnJ_f)\right],
\nn\\ \label{3.29} \er
with the same one-body matrix elements \rf{3.28}.

The unperturbed and perturbed transition strengths
are defined, respectively, as
\be
S^0_\mu(pnJ)=|{\Lambda}_\mu(pnJ)|^2
\label{3.30} \ee
and
\be
S_\mu(J_f)=|\Bra{J_f, Z+\mu,N-\mu}{\sf O}_{\sf J}\Ket{0^+}|^2
 \label{3.31} \ee
One might be  particularly interested in the Gamow-Teller (GT)
and Fermi (F) $\b$-decay strengths (B-values), in which
case \rf{3.30} and \rf{3.31} are evaluated
for the operators ${\widetilde{\sf O}}_{\sf J}$, which don't
contain the radial form factors $j_{\sf J}(\rho)$.
That is, ${\widetilde{\sf O}}_{0}=1$,
and ${\widetilde{\sf O}}_{1}=\mbs$, for F and GT operators,
respectively. We denote the B-values as $\widetilde{S}_\mu(pnJ)$
and $\widetilde{\S}_\mu(J_f)$.
Occasionally one also might want to calculate the  the energy
distribution of the last one, \ie
\br
\widetilde{S}_\mu(J_f,E)&=&\frac{\eta}{\pi}\sum_f
\frac{\widetilde{S}_\mu(J_f)}{\eta^2+(E-\omega_{J_f})^2},
\label{3.32}\er
where one usually takes $\eta=1$ MeV.

\section{Computer program and user's manual}

The QRAP code evaluates the electron neutrino-nucleus interaction
described by equation \rf{2.1} (IREAC=1 for NS, IREAC=2 for AS) and
\rf{2.2} (IREAC=0 for MC). The  processes,  from the ground state
of the even-even father nucleus
$(Z, N)$ to the excited states with
spin (ISPIN) and parity (IPARI) in the odd-odd daughter nucleus
$(Z\pm1, N\mp1)$,  are calculated by
using the QRPA model (IQP=0) or the PQRPA (IQP=1).
These options must be setup in the input data file,
\textbf{qrapin.dat}, which  is supplemented  with two included files:

(a) \textbf{sp.inc}, containing the dimensions of single-particle
quantum numbers, occupation probabilities, quasiparticle quantities
and strength amplitudes for allowed transitions;

(b) \textbf{conf.inc}, which has the dimensions for the quasiparticle
state configurations, the hamiltonian  matrices ($A, B$) or (${\cal A},
{\cal B}$) which are  diagonalized, the forward and backward amplitudes,
and the eigenvalues.

There are two  input files: 1) \textbf{qrapout.dat}, where is listed the
output file that shows the neutrino/antineutrino ($\nu/\bar{\nu}$) cross
section, as a
function of the incident neutrino, or the muon capture rate, for each
state of a given nuclear spin in  the daughter nucleus,  and 2) the above
mentioned \textbf{qrapin.dat}, which contains:  a) the quantum numbers
of all single-particle state (sps), and the corresponding single particle
energies (s.p.e.),
b) the mass and the proton number of the parent nucleus, c) the
neutron and proton pairing strengths for the BCS approximation,
d) the particle-particle, and particle-hole strengths of the residual
interaction, e) the position of the Fermi level, and the experimental
gap for neutrons and protons,  and f) the $Q$-value for the
$\nu/\bar{\nu}$ scattering.

There are three default output files. Two of them, \textbf{AUXI.OUT} and
\textbf{OUT.OUT}, contain the results of the nuclear structure
model, whereas the results for the weak processes appear in the file
created by \textbf{qrapout.dat}. For example, if one is interesting
in the multipole $J^\pi_f=1^+$ with a single-particle space of
six levels in $^{12}$C (``{\it set 1}''), we can introduce in
\textbf{qrapout.dat} the file names \textbf{QNC.out}
(\textbf{PNC.out}) for neutrino capture, \textbf{QAC.out}
(\textbf{PAC.out}) for antineutrino capture, \textbf{QMC.out}
(\textbf{PMC.out}) for muon capture, using the QRPA (PQRPA) model.
The auxiliary output files AUXI.OUT and OUT.OUT are relabeled to
(\textbf{QAUXI.OUT,~QOUT.OUT}) and (\textbf{PAUXI.OUT, POUT.OUT})
for QRPA and PQRPA respectively.

All just  mentioned outputs are included as examples.
The following units are employed: i) $10^{-42}$cm$^2$,
for  neutrino or antineutrino-nucleus cross sections,
ii) $10^{4}$s$^{-1}$, for muon capture rates,
and iii) MeV,  for  energies.

\subsection{Reading the data}

There are three sets of input data in \textbf{qrapin.dat}
separated in modules labeled as: \textit{*Data set 1} for a
single-particle space of six levels in $^{12}$C ($0, 1$, and
$2~\hbar\omega$ oscillator shells), \textit{*Data set 2} for a
single-particle space of ten levels in $^{12}$C ($0, 1, 2$, and
$3~\hbar\omega$ oscillator shells), and \textit{*Data set 3} for
single-particle space of 12 levels in $^{56}$Fe ($2, 3$, and
$4~\hbar\omega$ oscillator shells).

For each one of these input data, the number of sps
represents the available space where one
wants to solve the BCS~(or PBCS) problem given by equations \rf{3.17}
and \rf{3.18}~(\rf{3.1}-\rf{3.2}). It contains the necessary number
of harmonic oscillator shells leading to a smooth smearing of the
Fermi's surface. The Fermi level with the neighboring levels
constitute the active shell for the mentioned
smearing. For example, in $^{12}$C (ground
state with $J=0^+$) the active shell is composed by the
$1p_{3/2}$ and $1p_{1/2}$ levels. According to the
single-particle shell model the sps  filled up
the $1p_{3/2}$ orbital, and the nucleons can be  promoted to the
$1p_{1/2}$, creating a particle state in $1p_{1/2}$ and a hole state
in $1p_{3/2}$.  This scheme describes the first particle-hole (ph)
excitation on $^{12}$C in order to obtain the $^{12}$N or
$^{12}$B ground state with $J=1^+$, by promoting a proton or
a neutron, respectively.

Let us show, as example, a data input of six sps for $^{12}$C:
\textit{*Data set 1}. The rows starting with a symbol ``*''
are not read as input and just serve  to remind the user
on the meaning of the physical quantities. Taking out the
comments ``*'' in the first lines of this file, we have

{\small
~~1~~~+1~~~~0.0~~~~1~~~~1\\
~~06~~~~~~~06~~~~~0~~~~~~~1~~~~~~~~1~~~~~~~~0 \\
101 -20.09 112 -6.02 111 -0.29 123  3.07 201  3.85 122  7.18 \\
101 -18.19 112 -3.17 111 ~2.79 123  5.73 201  6.06 122  9.36 \\
~~~12~~~06~~~28.80~~~~~~28.85~~~~~30.0~~~~50.0~~~~27.0~~~64.0\\
~~~~~~~~~2~~~~~~~~~1~~~~~~~~~6~~~~~~~~~6~~~1.00~~~~~6.88 \\
~~~~~~~~~2~~~~~~~~~1~~~~~~~~~6~~~~~~~~~6~~~1.00~~~~~7.00 \\
~~17.338\\}

\textbf{First line:} Nuclear spin (ISPIN=1) of the
daughter nucleus, parity (IPARI=+1),
coupling strength of particle-particle channel ($t=0.0$) (see
definition below), index of neutrino reaction (IREAC=1) and
the index (IQP=1) to solve the PQRPA problem.

\textbf{Second line:} Number of  neutron sps (NSN=06),
number of proton sps (NSP=06), index to solve the QRPA
equation in the Tamm-Damcoff approximation (ITD: 0 no, 1 yes), index
to print the matrix elements of the nuclear hamiltonian to be
diagonalized (MAPR: 0 no, 1 yes), index to solve the BCS equation
with the self-energy term (IFMU: 0 no, 1 yes), index to
make the QRPA matrix with the branch ($Z+1,N-1$)
($\mu=1$) or ($Z-1,N+1$) ($\mu=-1$) of PQRPA
(IPRO: 0 ($\mu=1$), 1 ($\mu=-1$)).

\textbf{Third}  and \textbf{Fourth lines:} quantum numbers and the s.p.e.
for  each  neutron and proton sps, respectively.
They are represented in the same way as in the shell
model scheme, with their respective quantum numbers
($n~\ell~(j+\frac{1}{2}$)). For example, $101 \go 1s_{1/2}$ where $1$
is principal quantum number, corresponding to the first harmonic
oscillator level ($n$), $0$ corresponds to the orbital  angular
momentum $\ell \equiv s$ and the last number $1 \equiv
j+\frac{1}{2}=\frac{1}{2}+\frac{1}{2}$.
Table \ref{tab:1} shows the notation and the corresponding
quantum numbers, as well as the PBCS quasiparticle energies
$e_j^N$, and $e_j^Z$, defined
by \rf{3.10}.
%%%%%%%%%%%%%%%%%%%%%%%%%%%%%%%%%%%%%%%%%%%%%%%%%%%%%%%%%%%%%%%%%%%%%%%
\begin{table}[h]
\caption{Notation for the quantum numbers, the resulting
quasipartice energies $e_j^N$ for neutrons and $e_j^Z$ for
protons, and the pairing strength $v_{s}^{pair}$   within the PBCS.
The energies are given in units of MeV, and $v_{{s}}^{{pair}}$ is
dimensionless.}
\begin{center}
\label{tab:1}
\newcommand{\cc}[1]{\multicolumn{1}{c}{#1}}
\renewcommand{\tabcolsep}{0.5 pc}
\bigskip
\begin{tabular}{c c c c c |r r}%\hline
Notation&Shell&$n$&$\ell$&$j+1/2$&$e_j^N$&$e_j^Z$
\\ \hline
101&$1s_{1/2}$ &1&0&1&$-20.09$&$-18.19$
\\
112&$1p_{3/2}$ &1&1&2&$-6.02$  &$-3.17$
\\
111&$1p_{1/2}$ &1&1&1&$-0.29$ &$2.79$
\\
123&$1d_{5/2}$ &1&2&3&$3.07$  &$5.73$
\\
201&$2s_{1/2}$ &1&0&1&$3.85$  &$6.06$
\\
122&$1d_{3/2}$ &1&2&2&$7.18$  &$9.36$
\\ \hline
$v_{{s} }^{{pair}}$&&&&&$28.80$&$28.85$
\\\hline
\end{tabular}\end{center}\end{table}
%%%%%%%%%%%%%%%%%%%%%%%%%%%%%%%%%%%%%%%%%%%%%%%%%%%%%%%%%%%%%%%%%%%%%%%

\textbf{Fifth line:} Mass number $A$ (IAM=12), proton
number $Z$ (IZ=6), and the following coupling constants:
1)  neutrons and proton pairing $v_{{s}}^{{pairN}}$ (VspairN=28.80), and
$v_{{s} }^{{pairP}}$ (VspairP=28.85),
2) singlet and triplet particle-particle ($pp$) $v_{{s} }^{{pp}}$ (VsPP=30),
and $v_{{t}}^{{pp}}$ (VtPP=50), and iii)  singlet and triplet
particle-hole ($ph$) $v_{{s} }^{{ph}}$ (VsPH=27), $v_{{t} }^{{ph}}$ (VtPH=64).

\textbf{Sixth } and \textbf{Seventh lines:} Position of the Fermi level
(LEVEL=2), initial (IIQ=1) and final (IFQ=6) states for which the BCS
equations must be solved, number of particles interacting (NPIQ=6) in
these levels, and the experimental gap (DELTAQ=6.88 or 7.00) defined
below in equation \rf{4.2}; for neutrons and protons, fifth
and sixth lines respectively.

\textbf{Eighth line:} $Q$-value minus the lepton mass
for $\nu/\bar{\nu}$ scattering
(EGS=17.338 for $^{12}$N \cite{Ajz85}). It can be fixed  as being the
energy of the ground state in the  daughter nucleus. The lepton mass
must be added to EGS to obtain the $Q$-value for the reaction.

%%%%%%%%%%%%%%%%%%%%%%%%%%%%%%%%%%%%%%%%%%%%%%%%%%%%%%%%%%%%%%%%%%%%%%%
\begin{table}[t]
\caption{Spin and parity for the one-quasiparticle space used in the
input for $^{12}$C.}
\begin{center}
\label{tab:3}
\newcommand{\cc}[1]{\multicolumn{1}{c}{#1}}
\renewcommand{\tabcolsep}{0.2 pc}
\bigskip
\begin{tabular}{c|c c c c c c}
&$1s_{1/2}$&$1p_{3/2}$&$1p_{1/2}$&$2s_{1/2}$&$1d_{5/2}$&$1d_{3/2}$
\\ \hline
$1s_{1/2}$ &$0^+,1^+$&&&&
\\
$1p_{3/2}$ &$1^-,2^-$&$0^+,1^+$&&&
\\
                    &&$2^+,3^+$&&&
\\
$1p_{1/2}$ &$0^-,1^-$&$1^+,2^+$&$0^+,1^+$&&
\\
$2s_{1/2}$ &$0^+,1^+$&$1^-,2^-$&$0^-,1^-$&$0^+,1^+$&
\\
$1d_{5/2}$ &$2^+,3^+$&$1^-,2^-$&$2^-,3^-$&$2^+,3^+$&$0^+,1^+,2^+$
\\
&&$3^-,4^-$&&&$3^+,4^+,5^+$
\\
$1d_{3/2}$ &$1^+,2^+$&$0^-,1^-$&$1^-,2^-$&$1^+,2^+$&$1^+,2^+$&$0^+,1^+$
\\
                    &&$2^-,3^-$&         &         &$3^+,4^+$&$2^+,3^+$
\end{tabular}\end{center}\end{table}

\subsection{Running the code}

As first step the QRAP solves the BCS problem. In this case,
one needs to adjust
the pairing strength to reproduce the experimental pairing gap.

Next, one can solve the PBCS problem or directly the QRPA if the option
IQP=0 was selected. If IQP=1 then the PQRPA equations are solved.
It means that QRAP firstly calculates the nuclear matrix elements
in the QRPA or PQRPA by selecting the option IQP=0 or 1, appropriately.
The option for which type of weak interaction process one wants
to evaluate is adopted with IREAC in the input data.
We recommend first to adjust the pairing strength as it is explained
below.  After this it is convenient to
fit the parameters of the residual interaction using the option
IREAC=3 for the muon capture rate because this calculation is
fast. Physically you can check quickly how good is your choice  of
parameters because the values for inclusive muon-capture rate, and GT
B-values are available in the literature
(see for example Refs.~\cite{Suz87,Cho93}).

For the residual interaction the code assumes  a delta force,
\br
V&=&-4 \pi \left(v_sP_s+v_tP_t\right)
\delta(r),
\label{4.1}\er
which  has been used extensively in the literature
\cite{Hir90a,Hir90b,Krm92} to describe single and double beta
decays.

Next, we explain how the parameters of the interaction are
adjusted using for example the input data for six levels
in $^{12}$C. The results are presented in output file
\textbf{OUT.OUT}.

$\bullet$ \underline{Adjusting the gap $\Delta_k$:}

The parameters $v_{{s} }^{pairN}$ and $v_{{s} }^{pairP}$ are
adjusted to reproduce the experimental gap $\Delta^N$ for neutrons,
and $\Delta^Z$ for protons, by  solving the BCS equations  \rf{3.17}
and \rf{3.18}  in a self-consistent way.
The experimental gaps, according~\cite[Eq. 2.96]{Boh69}, are:
\br
\Delta^N=-{1\over2}\left\{{\cal B}(Z,N-1)-2{\cal B}(Z,N)+{\cal
B}(Z,N+1)\right\},
\nn\\
\Delta^Z=-{1\over2}\left\{{\cal B}(Z-1,N)-2{\cal B}(Z,N)
+{\cal B}(Z+1,N)\right\},
\nn\\
\label{4.2}\er
where ${\cal B}(N,Z)$ is the binding energy of the
even-even nucleus $(Z,N)$. This is the most common fit (Fit1) used in
several works with standard  QRPA \cite{Krm92,Krm93,Krm94}. In this
case, the $\Delta^{N(Z)}$ must be equal or approximately equal to the
energy $\Delta_{j_k=FL}^{N(Z)}$ of the corresponding to the Fermi
level (FL). To solve the set of PBCS coupled equations
\rf{3.1}-\rf{3.4} for $u_k$ and $v_k$ it is recommended to obtain
first the solutions for the BCS problem, as
these probability occupations are use as input for the PBCS case.
The PBCS coupled nonlinear equations are solved consistently
with Powell Hybrid method using subroutine HYBRD \cite{Arg80}.

The results
of the BCS or PBCS problem are shown as tables in the first lines of
\textbf{OUT.OUT} for neutrons and protons, respectively. The
quantities defined by \rf{3.10} and \rf{3.11} are presented there.  In
particular, the projected quasiparticle energy defined in \rf{3.10} are
\begin{eqnarray}
&{\rm PROYSP}=&
\left\{
\begin{array}{l}
\rm{E}(+)=\varepsilon^{K}_k,\;\;\mbox{with}\;k\;
\mbox{above Fermi level},
\\
\rm{E}(-)=\varepsilon^{K-2}_k,\;\;\mbox{with}\;k\;
\mbox{below Fermi level,}
\\\end{array}\right.
\nn\\
\label{4.3}
\end{eqnarray}
which means that $\rm{E}(+)$ corresponds to a particle state,
and  $\rm{E}(-)$ to a hole state. The values of
$\Delta_{j_k}^{N(Z)}$ are shown in the ninth
column of the table labeled as CONFIGURATION SPACE.
This Fit1 comes from the fact that the experimental energy
difference between the states that lie just above (p state) and
just above (h state) the FL is approximately twice
the experimental gap, \ie
\be
E_p^{K}-E_h^{K} \simeq 2 \Delta^{K}
\label{4.4}\ee
 for   $K=N$ or $Z$.

There is another fitting procedure for the pairing gap that is
called by Fit2. In Fit2, all the s.p.e.  $e_j^{N(Z)}$ from
Table I are varied with a $\chi^2$ search  to account for the
experimental spectra $E_j$:
\brn &&\epsilon_{k}^{Z(N)} \go
E_p^{Z(N)}\equiv {\rm E}(+),~~~  \mbox{for a particle state,}
\nn\\
&&\epsilon_{k}^{Z-2(N-2)} \go E_h^{Z(N)}\equiv {\rm E}(-) ,~~~
\mbox{for  a hole state}.\ern
In  Fit 2, the Eq. \rf{4.4}
is automatically satisfied. This procedure was employed  to obtain
the $e_j$ spectra shown in  ~\cite[Table III]{Krm05},  whereas
the $e_j$ for the reduced space of six levels in the present example
are shown in Table \ref{tab:1}. These  s.p.e. are used in
input data {\it Data set 1}.

To make the calculations as simple as possible the Fit1
procedure is the usual choice, with the $e_j$ spectra obtained
either from a harmonic oscillator or from a Wood-Saxon potential,
and by varying  the coupling $v_{{s} }^{pairN}$ and $v_{{s} }^{pairP}$
to satisfy the condition $\Delta_{j_k=FL}^{N(Z)} \approx \Delta^{N(Z)}$.
%%%%%%%%%%%%%%%%%%%
\begin{figure}[t]
\vspace{-2cm}
\begin{center}
{\includegraphics[width=9.cm,height=12.cm]{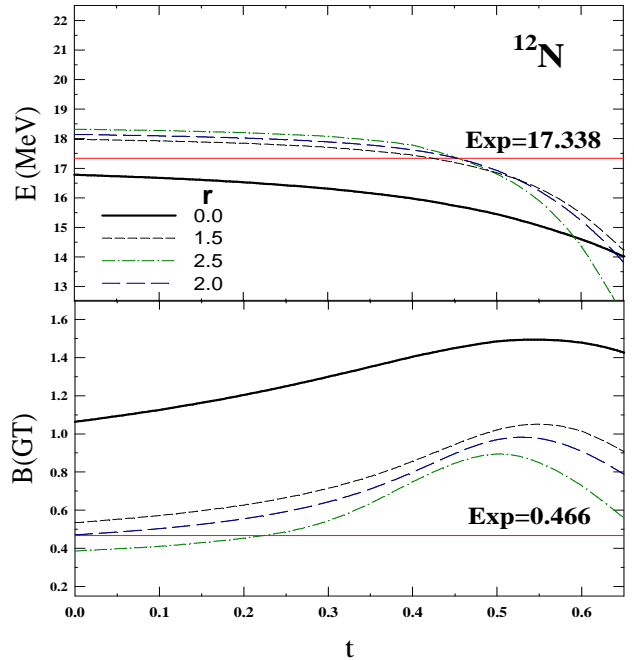}}
\end{center}
\vspace{-2.cm} \caption{ Ground state energy
E, and $B(GT)$-value in $^{12}$N for different
couplings $r$ in the $ph$-channel, as a function  of the
$pp$-channel coupling $t$. The experimental
data ~\cite{Ajz85,Al78,Mill72} are also shown.  The value $r=2$
corresponds to $v^{ph}_s=27$ a nd $v^{ph}_t=64$
(Case PII of \cite{Krm05}). The \textit{Data set 2} was employed.}%
\label{fig1}\end{figure}
%%%%%%%%%%%%%%%%%%%

For $^{56}$Fe the input data is
called \textit{*Data set 3}. The s.p.e  of the
active $3\hbar\w$  shell were taken  from the experimental energies
of $^{56}$Ni, and the $2\hbar\w$ and $4\hbar\w$ shell energies were
taken from the harmonic oscillator energies with $\hbar\w/{\rm
MeV}=~45~A^{1/3}-25~A^{2/3}$. Fit1 was employed to adjust the
experimental $\Delta^{N(Z)}$ for $^{56}$Fe.

%%%%%%%%%%%%%%%%%%%
\begin{figure}[t]
\begin{center}
{\includegraphics[width=7cm,height=8.cm]{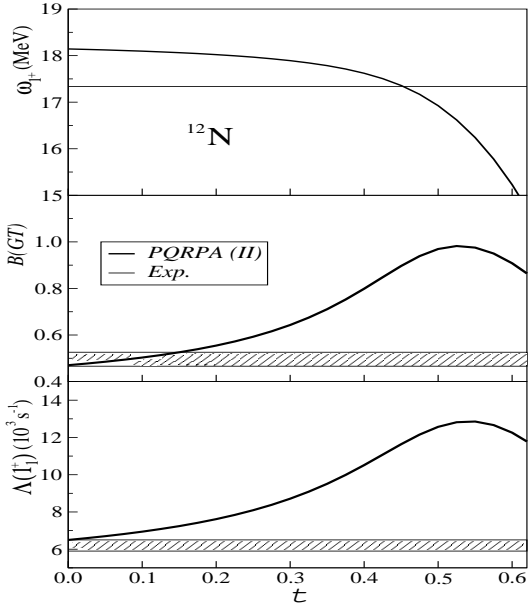}}
%{\includegraphics[width=9.cm,height=12.cm]{FIG6thesis.eps}}
\end{center}
\caption{Results from PQRPA  calculations obtained  in
Ref.~\cite{Krm05}, as a function of the $pp$ parameter $t$, compared
with the experimental data  taken from
Refs.~\cite{Ajz85,Al78,Mill72}, for: (i) ground state energy in
$^{12}$N (upper panel), (ii) $B(GT)$-value for the $\beta$
transition in $^{12}$N (middle panel), and (iii)
exclusive muon capture rate  $\Lambda(1^+_1)$ in $^{12}$B
(lower panel). The parameters of the $ph$ channel for the
$\delta$-interaction are $v^{ph}_s=27$ and $v^{ph}_t=64$.}%
\label{fig2}\end{figure}

$\bullet$ \underline{Adjusting the particle-hole couplings $r$
and $p$}

In the particle-hole matrix element $F$, defined in
Eq.~\rf{3.5}, the couplings $v_s$ and $v_t$ appear as linear combinations
$v_s+v_t$ and $3v_t-v_s$. Therefore,  it is convenient to introduce
the dimensionless parameters
\br
r=\frac{v_s^{ph}+v_t^{ph}}{2v_s^{pair}},\hspace{.5cm}p
=\frac{3v_t^{ph}-v_s^{ph}}{2v_s^{pair}},
\label{4.5}\er
where
\be
v_s^{pair}=\frac{v_s^{pairN}+v_s^{pairP}}{2}.
\label{4.6}\ee
Moreover, to start we can use $v^{ph}_s=27$ and
$v^{ph}_t=64$~ (in units of MeV.fm$^3$). These values are
inferred  from the systematic study of energetics of the GT resonances
done by Nakayama \etal~\cite{Nak82} (see also Ref.~\cite{Cas87}),
and  have been extensively used in the QRPA calculations of
the $\b\b$-decay in  $^{48}$Ca~\cite{Hir90,Krm94,Bar98}.
Moreover, it makes sense  to take the  singlet ph coupling to be
equal to $v^{pair}_s$, obtained from the proton and neutron gap
equations, \ie
\be
v_s^{ph}=v_s^{pair}.
\label{4.7}\ee

$\bullet$ \underline{Adjusting the
particle-particle couplings $s$ and $t$}

Here also it is  convenient to  normalize to $v_s^{pair}$ the coupling
constants $v^{pp}_s$ and $v^{pp}_t$ that appear in the $pp$ matrix
elements $G$ in Eq.~\rf{3.5}, and  correspond, respectively  to
the channels $(T=1, S=0)$ and $(S=1, T=0)$, \ie
\be
s=\frac{v_s^{pp}}{v_s^{pair}},\hspace{0.5cm}t=
\frac{v_t^{pp}}{v_s^{pair}}.
\label{4.8}\ee
For nuclei with $N > Z$ the $pp$-couplings are fixed on the basis
of the SU(4) and isospin  symmetry, as $v^{pp}_s \equiv v^{pair}_s$,
and $v^{pp}_t \gsim v^{pp}_s$~\cite{Hir90,Krm94}.  However,
in Ref~\cite{Krm02} it was shown  that this
parametrization  might not be suitable for $N=Z$. In fact, the best
agreement with data in $^{12}$C was obtained when the $pp$-channel
is totally switched off, \ie $v^{pp}_s\equiv v^{pp}_t=0$, and three
different set of values for the ph-coupling strengths were used.
These conditions are related with $s=t$ for $^{12}$C ($N=Z$), and
with $s=1$ and $t$ variable in nuclei with $N>Z$.
For  $^{56}$Fe  were adopted the values $s=1$ and
$t=0$. In the code QRAP the following conditions  are
standard: (i) $s=t$ with $t$ as a variable parameter for $N=Z$; and
(ii) $s=1$, and $t$ as a variable parameter for $N>Z$, \ie  the
residual interaction is defined as a function of two adjustable
parameters $v^{ph}_t$ and $t$.

Several experimental data are  available in the
literature that can be used for fixing the residual interaction
coupling constants, such as: ground state energies of daughter nuclei,
$B(GT)$-values for the $\beta^+$ or $\beta^-$ decay, and partial
muon capture rate ~\cite{Ajz85,Al78,Mill72}.

One  can use the reduced space of six levels to identify in the
output file, the quantities shown in Figure~\ref{fig1}. The results
 for   three values of $t$  are shown in Table~\ref{tab:3_2}.
%%%%%%%%%%%%%%%%%%%%%%%%%%%%%%%%%%%%%%%%%%%%%%%%%%%%%%%%%%%%%%%%%%%%%%%
\begin{table}[h]
\caption{Evolution of  ground state energy, $B(GT)$ and exclusive muon
capture rate in $^{12}$C, as function of the $pp$-channel parameter $t$.
With [a] and [b] we denote, respectively,  the output files "AUXI.OUT"
and "PMC.out", with $\omega_{\mu}(1^+_{f})$ is in units of MeV,
${S}_{\mu}(1^+_{f})$ is dimensionless and
($\Lambda(1^+_{1})$, $\Lambda$) is in units of $10^4$ s$^{-1}$.
The parameters for the $ph$-channel are: $v_s^{PH}=27, v_t^{PH}=64$.}
\begin{center}
\label{tab:3_2}
\newcommand{\cc}[1]{\multicolumn{1}{c}{#1}}
\renewcommand{\tabcolsep}{0.1 pc}
\renewcommand{\arraystretch}{1.2} % enlarge line spacing
\bigskip
\begin{tabular}{c|c|l| c c c c}%\hline
&&&&$t$&
\\
&State~[File]&Observable &0.0&0.3&0.6& Exp.
\\ \hline
$^{12}$N
&$16$~[a]  &~$\omega_{+1}(1^+_{16})$ &18.319&17.951&14.970&17.34~\cite{Ajz85}
\\
&$16$~[a]  &~${S}_{+1}(1^+_{16})$    & 0.496&0.696&0.840&0.466~\cite{Al78}
\\ \hline
$^{12}$B
&$1$~[a]   &~$\omega_{-1}(1^+_{16})$ &12.528&12.126&9.202&13.36~\cite{Ajz85}
\\
&$1$~[a]   &~${S}_{-1}(1^+_{16})$    & 0.502&0.693&0.837&0.526~\cite{Al78}
\\
&$16$~[b]  &~$\Lambda_{+1}(1^+_{16})$& 0.689&0.936&1.119&$0.62(3)$~\cite{Mill72}
\\
& ~[b]&~$\sum_f\Lambda_{+1}(1^+_{f})$& 1.722&1.537&1.183&
\\ \hline \hline
\end{tabular}\end{center}\end{table}
%%%%%%%%%%%%%%%%%%%%%%%%%%%%%%%%%%%%%%%%%%%%%%%%%%%%%%%%%%%%%%%%%%%%%%

The values of $\omega_{\mu}(1^+_{f})$, and  ${S}_{\mu}(1^+_{f})$
in $^{12}$N and $^{12}$B can be found in the output file {\bf AUXI.OUT}.
In the present case
the  largest value of index $f$ is $f_{max}$=16.
Both set of states, with $\mu=+1$, and  $\mu=-1$,  are ordered from
highest to lowest energies.
In the PQRPA, the most collective ones are that of the
corresponding ground states:
$\ket{1^+_{F=16}}$ in $^{12}$N
(and $\ket{1^+_{F=17}}$  in $^{12}$B) although there also are
significant strengths in the states $F=7,11$, and $14$.
In QRPA, the ground state is in $\ket{1^+_{F=16}}$
for both  $^{12}$N and $^{12}$B. These wave functions are presented
below. For the PQRPA case, we also
show  the  unperturbed energies $\omega^0_{\mu}(pn1^+)$ (which are not
ordered),  and the corresponding single-particle GT strengths
${S}^0_{\mu}(pn1^+)$, given respectively, by \rf{3.8}, and
\rf{3.30} for the GT operator $\mbs$. The largest ones are
${S}^0_{+1}(1p_{1/2}^\pi,1p_{3/2}^\nu;1^+)$,
and ${S}^0_{-1}(1p_{3/2}^\pi,1p_{1/2}^\nu;1^+)$, which in
the particle-hole limit correspond to excitations
$1p_{3/2}^\nu\go 1p_{1/2}^\pi$, and  $1p_{3/2}^\pi\go 1p_{1/2}^\nu$.
For spins and parities  $J^\pi_f\ne1^+$, or $0^+$, in {\bf AUXI.OUT}
are shown the energies $\omega_\mu$, but not the strengths $S_\mu$.

The results for the eigenvalue problem  are displayed in the output
file \textbf{OUT.OUT}. For the option MAPR= 1 are printed out the matrix
elements $({\cal A, B})$ Eq.~\rf{3.25} for the QRPA,
or $(A_{\mu=1}, B)$ eq.~\rf{3.7} for PQRPA.
The nuclear wave functions ($X(pn;J_{F}),Y(pn;J_{F})$) are grouped to
four, with the index $F$, defined in \rf{3.12}, going from 1
to $f_{max}$ in the QRPA case, and from 1  to $2f_{max}$ in the
PQRPA case. To make easy reading together with each set of wave
functions are  also printed: the value of $f$, the two quasiparticle
configurations ($p$ and $n$), and the unperturbed and perturbed energies.

Recalling  Eqs.~\rf{3.8},  \rf{3.13}, and  \rf{3.14} for the
energies, one discovers without difficulty that within PQRPA:

1) The ground state in $^{12}$N, with
energy $\omega_{+1}(1^+)=18.319$ MeV, has $f=F=16$, and that
its wave function is:
\br
\ket{^{12}{\rm N}}&=&0.963\ket{1p^\pi_{3/2}1p^\nu_{1/2},1^+}
+0.232\ket{1p^\pi_{3/2}1p^\nu_{3/2},1^+}
\nn\\
&+&0.122\ket{1p^\pi_{1/2}1p^\nu_{3/2},1^+}
+0.105\ket{1p^\pi_{1/2}1p^\nu_{1/2},1^+}
\nn\\
&+&\cdots\label{4.9}\er

2) The ground state in $^{12}$B, with
energy $\omega_{-1}(1^+)=12.528$ MeV, has $f=16,~F=17$,
and that  its wave function is:
\br
\ket{^{12}{\rm B}}&=&-0.971\ket{1p^\pi_{1/2}1p^\nu_{3/2},1^+}
+0.204\ket{1p^\pi_{3/2}1p^\nu_{3/2},1^+}
\nn\\
&-&0.125\ket{1p^\pi_{3/2}1p^\nu_{1/2},1^+}
+0.090\ket{1p^\pi_{1/2}1p^\nu_{1/2},1^+}
\nn\\
&+&\cdots\label{4.10}\er

One proceeds  in a similar way for the QRPA output, with energies
now given by Eqs. \rf{3.23} and  \rf{3.24}. Now,  the ground state
energies  in $^{12}$B, and $^{12}$N, are, respectively,
$\omega_{-1}(1^+)=12.437$ MeV, and $\omega_{+1}(1^+)=17.992$ MeV,
while the  wave function for both nuclei is: \br
\ket{1^+_{16}}&=&-0.272\ket{1p^\pi_{3/2}1p^\nu_{1/2},1^+}
-0.759\ket{1p^\pi_{3/2}1p^\nu_{3/2},1^+}
\nn\\
&+&0.356\ket{1p^\pi_{1/2}1p^\nu_{3/2},1^+}
-0.472\ket{1p^\pi_{1/2}1p^\nu_{1/2},1^+}
\nn\\
&+&\cdots\label{4.11}\er

From the comparison of the wave functions \rf{4.9}, and \rf{4.11}
it can be easily figure out why Volpe \etal ~\cite{Vol00} called
attention to {\em ``difficulties in choosing the ground state of
$^{12}$N, because the lowest state is not the most collective
one"} when the QRPA is used. This is an important issue that
clearly gives you an idea about the need for the number
projection. In fact, as seen from \rf{4.9}, and \rf{4.11}, the
PQRPA yields the correct one-particle-one-hole (1p1h) limits
$1p^\pi_{3/2}\go 1p^\nu_{1/2}$ and $1p^\nu_{3/2}\go 1p^\pi_{1/2}$,
for  $^{12}$N, and $^{12}$B ground states, respectively. All
remaining configurations comes from the higher order 2p2h, and
3p3h excitations. Contrary, the QRPA  state \rf{4.11} is
dominantly the two-hole excitation
$[(1p^\pi_{3/2})^{-1},(1p^\nu_{3/2})^{-1}]$, which corresponds to
the ground state of $^{10}$B. This should not be a surprise, as we
know that the proton-neutron QRPA states are the same for all four
nuclei $^{12}$N, $^{10}$B, $^{14}$N, and  $^{12}$B. More details
on this question can be found in \cite[Fig. 3]{Krm05}. The 1p1h
amplitudes$[(1p^\pi_{3/2})^{-1},1p^\nu_{1/2}]$, and
$[(1p^\nu_{3/2})^{-1},(1p^\pi_{1/2})]$  are dominantly present in
the QRPA states $f=13$, and   $f=15$,~\ie \br
\ket{1^+_{13}}&=&-0.476\ket{1p^\pi_{3/2}1p^\nu_{1/2},1^+}
+0.437\ket{1p^\pi_{3/2}1p^\nu_{3/2},1^+}
\nn\\
&+&0.441\ket{1p^\pi_{1/2}1p^\nu_{3/2},1^+}
-0.096\ket{1p^\pi_{1/2}1p^\nu_{1/2},1^+}
\nn\\
&+&\cdots\nn\\
\ket{1^+_{15}}&=&0.703\ket{1p^\pi_{3/2}1p^\nu_{1/2},1^+}
+0.708\ket{1p^\pi_{1/2}1p^\nu_{3/2},1^+}
\nn\\
&+&\cdots\label{4.12}\er

The  wave functions displayed above clearly evidence the superiority of the PQRPA on the QRPA.

$\bullet$ \underline{Output for the $\nu$-nucleus processes}

The output of the results for the  weak processes is selected
according to the value of \textbf{IREAC}:

\textbf{IREAC=0} prints the results for the muon capture rate in the
file (QMC.out or PMC.out).
For  $J^\pi=0^+$ or $J^\pi=1^+$ are shown in in this output
file the folded strengths $\widetilde{S}_\mu(J_f,E)$ (S\^{}TILDE)
defined by \rf{3.32}, where `ENERGY' represents $E$.
The partial capture rate for each state $f$, the perturbed
energy $\omega_{\mu=-1}$, and the strength $S_{\mu=-1}(J_f)$
(if $J^\pi=0^+$, or $J^\pi=1^+$) are shown in the table labeled
CAPTURE RATE. The total capture for the evaluated spin  $J^\pi$ is
presented in the last line.

\textbf{IREAC=1}  or \textbf{IREAC=2}  prints the results for the
neutrino or antineutrino cross sections in the files
(QNC.out/PNC.out) or (QAC.out/PAC.out). We repeat in this output
the folded strengths $\widetilde{S}_\mu(J_f,E)$ \rf{3.32}
for $J^\pi=0^+$  or $J^\pi=1^+$.
The cross sections (SIGMA(Enu)) are calculated as a function
of the neutrino energy (Enu) for each nuclear spin
from $f=1$ to $f=f_{\rm max}$. The perturbed $\omega_{\mu=\pm 1}$
energies for the daughter nucleus are also shown according the
process related. The absolute value of maximum~($\cos\theta=-1$)
and minimum ($\cos\theta=1$) nuclear momentum transfer ($|{\rm k}|$)
in units of MeV/c  for  each energy are also printed.

\textit{Note:} The cross sections are printed up to a maximum
energy of 250 MeV. Depending upon the single particle space
employed, the cross sections, as a function of the neutrino energy,
should be restricted to lower energies. This issue
will be discussed and explained in details in a next work~\cite{Sam10}.
Anyway, the PQRPA cross sections obtained within the single particle
space provided as examples are well behaved up to
$E_{\nu/\bar{\nu}} <$ 100 MeV on averaged according to
\textit{*Data set 1}, \textit{*Data set 2} and \textit{*Data set 3}.
This interval of energies is important for  supernova neutrinos and
low-energy decay-at-rest neutrinos~\cite{Arm02} .

\section{Routines included with the code}

\textbf{QRPA} solves the $pn$-QRPA or $pn$-PQRPA charge-exchange
problem for a nuclear spin $J^\pi_f$ of the daughter odd-odd nucleus.

\textbf{SUAVE} calculates and prints the folded strength $J^\pi=0^+$
or $J^\pi=1^+$ given by Eq. \rf{3.32}, folding the
$\widetilde{S}_\mu(J_f)$ stenght with a Lorentzian function
with $\eta=1$ MeV.

\textbf{RMUONCAP} calculates the muon capture rate given
by formula \rf{2.29}.

\textbf{SIMPSN2} calculates the neutrino or antineutrino cross
sections as a function of the neutrino energy. This subroutine
uses the function {\bf G} to call the
subroutine \textbf{SECCION}, which evaluates the cross section
formula \rf{2.23} using the Gauss-Legendre N-point quadrature
formula~\cite{Num86} on the function {\bf F} to evaluate the angular
integration of the transition amplitude times $E_\ell$.

\textbf{MATRIXP} computes the matrix elements with the delta
residual interaction given in Ref.~\cite{Sup64} for the PQRPA. The
matrix elements were  modified according to  the projection
procedure shown in Eq. \rf{3.7}.

\textbf{MATRIX} computes the matrix elements with the delta residual
interaction given in Ref.~\cite{Sup64} for the QRPA. The matrix
elements are shown in Eqs.~\rf{3.25}.

\textbf{RPA} finds eigenvalues and eigenvectors for the QRPA or
PQRPA equations. It uses the subroutine \textbf{EIGRF} and other
related subroutines from the IMSL Library \cite{IMSL}
to orthonormalize  the eigenvectors.

\textbf{GAPII} solves the set of BCS coupled Eqs.
\rf{3.17} and \rf{3.18} to obtain the $v_k$ and $u_k$ for
neutrons and protons.

\textbf{CONFGT} builds up the  pn configurations for  a given
spin and parity.

\textbf{FAUX} evaluates the particle-particle matrix elements
$G(kkk'k';0)$ and $F(kkk'k';0)$, which are used to solve the gap
equations for neutrons and protons using the delta interaction.

\textbf{RADWF} computes harmonic oscillator radial wave functions.
It uses the additional subroutine \textbf{OSCILL} to evaluate the
radial coefficients.

\textbf{HYBRD} finds a zero of a system of $N$ nonlinear equations
in $N$ variables by a modification of the Powell Hybrid method. This
subroutine was provided by the Argonne National Laboratory
\cite{Arg80}. It uses the subroutine \textbf{FCN} to calculate the
PBCS nonlinear equations given by formula \rf{3.1}.

\textbf{FKPERMAT} evaluates the perturbed
 matrix elements for the weak %beta
decay operator,  according to
Eq.~\rf{3.27} for PQRPA, and Eq. \rf{3.29} for QRPA. The radial
part of the SPNME were defined in Ref.~\cite{Ike64}.This subroutine
uses the subroutine \textbf{ANGULARMATRIX} to calculate the
angular part of single-particle matrix elements defined in
Ref.~\cite{Bar98}, and shown in the Appendix A for the sake of
completeness.

There are other routines in the code that are shortly described
as follows.
\textbf{PRINMA} prints the matrix elements
$({\cal A}, {\cal B})$ for the QRPA, or $(A_\mu,B)$ for PQRPA,
 \textbf{SKIPCOM} is used to skip comments in the input file,
\textbf{UNPMOM3} evaluates the unperturbed projected matrix
elements for beta decay, \textbf{BETMAT2} is used to calculate
the single-particle matrix elements for beta decay, \textbf{PROENER}
calculates the quantities for the projected quasiparticles
energies in \rf{3.10}.

\section{Things to do}

1. Use the sample input {\it Data set 1} to obtain the results
presented in Table \ref{tab:3_2}.

2. Modify the input {\it Data set 1} by {\it Data set 2}, setting
all parameters of the residual interaction to zero. These values
correspond to BCS or PBCS approximation. Compare the folded
strength of {\it Data set 1} with {\it Data set 2} shown in
Fig. 4 of Ref.~\cite{Krm05}.

3. In Ref.~\cite{Sam07} the s.p.e. for neutrons were changed to
analyze the systematics of the paring strength in the odd carbon
isotopes. Change the s.p.e for neutrons in {\it Data set 2} and
reproduce the systematics shown in the level scheme of Fig. 2 and
the spectroscopic factors of Fig. 3 of Ref.~\cite{Sam07}.

4. Compare the  QRPA and PQRPA results for the exclusive
$\nu_e-^{12}$C cross section,  as a function of the neutrino
energy, with the DAR experimental data from Ref.~\cite{Ath97}.
Note that the QRPA result is not collective, and the addition of
other $1^+$ cross sections (for example, that of  states
\rf{4.12}) is required to get agreement with the experimental
value.

\section*{Acknowledgments}

This work was partially supported by the U.S. DOE grants
DE-FG02-08ER41533 and DE-FC02-07ER41457 (UNEDF, SciDAC-2), and the
Research Corporation. F.K. thanks the
 CONICET-Argentina for the financial support through PIP-6159 and PIP-0377. A.R.S. wish to express
their sincere thanks to C.A. Barbero and A.E. Mariano for help
received in programming the s.p. matrix elements.

\section*{Appendices}

\subsection{Single-particle nuclear matrix elements}

The elementary operators defined in Eq.  \rf{2.16} have the
reduced single-particle $pn$ matrix elements (RSPME) defined in
\cite{Bar98,Don79} (Recall that $\kappa=|\kb|$ and
$\pb=-i\mbn $).

For the RSMPE dependent
on the tensor product of spherical harmonic times
the nucleon velocity we have
\br
&\Bra{p,(l_p~\fot),j_p}
j_{\sf L}(\kappa r)[Y_{\sf L}(\hat{\rb})\otimes\mbn]_{{\sf J}}
\Ket{n,(l_n~\fot),j_n}=&
\nn \\
&{(-1)^{\sf 1+J+L}\over \sqrt{4\pi}}
\left[W_{\sf LJ}^{(-)}(pn) R_{\sf L}^{(-)}(pn;\kappa)
+W_{\sf LJ}^{(+)}(pn) R_{\sf L}^{(+)}(pn;\kappa)\right],&
\nn\\\label{A1}\er
with angular and radial parts, respectively:
\br
W_{\sf LJ}^{(\pm)}(pn)&=& \pm  (-1)^{l_p+j_n+{\sf J}+1/2}
\hat{\sf J}\hat{\sf L}\hat{l}_p\hat{j}_p\hat{j}_n
(l_n+\frac{1}{2}\mp\frac{1}{2})^{1/2}
\nn\\
&&(l_p {\sf L}|l_n\mp1) \sixj{l_p}{j_p}{\frac{1}{2}}{j_n}{l_n}{\sf J}
\sixj{\sf L}{\sf J}{1}{l_n}{l_n \mp 1}{l_p},
\nonumber\\
R_{\sf L}^{(\pm)}(pn;\kappa)&=&\int_0^\infty u_{n_p,l_p}(r)
\left(\frac{d}{dr}\pm \frac{2l_n+1\pm1}{2r} \right)
\nonumber\\
&&u_{n_n,l_n}(r)~ j_{\sf L}(\kappa r)~ r^2~ dr.
\label{A2}\er
We use here the angular coupling
$\ket{(\frac{1}{2},l)j}$, $\hat{J}\equiv\sqrt{2J+1}$
and $(l_p {\sf L}|l_n\mp1)$ is the short notation for the
Clebsh-Gordon coefficient $(l_p 0 {\sf L} 0|(l_n\mp1)~0)$.

For the scalar product of spin times
nucleon velocity, we have
\br
&\Bra{p,(l_p~\fot),j_p}
j_{\sf J}(\kappa r)Y_{\sf J}(\hat{\rb})(\mbs\cdot\mbn)
\Ket{n,(l_n~\fot),j_n}=&
\label{A3} \\
&{1\over \sqrt{4\pi}}\left[W_{\sf J}^{(-)}(pn)
R_{\sf J}^{(-)}(pn;\kappa) +W_{\sf J}^{(+)}(pn)
R_{\sf J}^{(+)}(pn;\kappa)\right],&
\nonumber\er
with the angular part
\br
&W_{\sf J}^{(\pm)}(pn)= \pm  (-1)^{l_n+j_n+{\sf J}+1/2}
\sqrt{6}\hat{\sf J}\hat{l}_p\hat{j}_p\hat{j}_n
(l_n+\frac{1}{2}\mp\frac{1}{2})^{1/2}&
\nn\\
&(l_p {\sf J}|l_n\mp1) \sixj{1}{\frac{1}{2}}
{\frac{1}{2}}{j_n}{l_n}{l_n \mp 1}
\sixj{l_n \mp 1}{j_n}{\frac{1}{2}}{j_p}{l_p}{\sf J},&
\label{A4}\er
being the radial part $R_{\sf J}^{(\pm)}(pn;\kappa)$
as in \rf{A2}.

The RSPME of the two operator independent of the nucleon
velocity are written below.
For the the spherical harmonic operator we have
\br
&\Bra{p,(l_p~\fot),j_p}
j_{\sf J}(\kappa r)Y_{\sf J}(\hat{\rb})
\Ket{n,(l_n~\fot),j_n}=&
\label{A5} \\
&{1\over \sqrt{4\pi}} W_{\sf J0}(pn) R_{\sf J}^{0}(pn;\kappa)&,
\nonumber\er
with the angular and radial parts, respectively:
\br
&&W_{\sf J0}(pn)=(-1)^{j_p-j_n} \hat{\sf J}
\hat{j}_p \hat{j}_n
\threej{j_p}{j_n}{\sf J}{1\over2}{-{1\over2}}{0},
\label{A6}\\
&&R_{\sf J}^0(pn;\kappa)=\int_0^\infty u_{n_p,l_p}(r)
u_{n_n,l_n}(r)~ j_{\sf J}(\kappa r)~ r^2~ dr.
\nn\er
Finally,
the RSMPE dependent of the tensor product of spherical harmonic
times the spin operator reads
\br
&\Bra{p,(l_p~\fot),j_p}
j_{\sf L}(\kappa r)
\left[Y_{\sf L}(\hat{\rb})\otimes{\mbs}\right]_{\sf J}
\Ket{n,(l_n~\fot),j_n}=&
\nn \\
&{(-1)^{\sf L+1+J}\over \sqrt{4\pi}} W_{\sf LJ}(pn)
R_{\sf L}^{0}(pn;\kappa),&
\label{A7}\er
where  the angular part is
\br
W_{\sf LJ}(pn)&=&(-1)^{l_p} \sqrt{6}
\hat{j}_p \hat{j}_n \hat{l}_p \hat{l}_n \hat{j}_p
\hat{\sf L} ~\hat{\sf J}
\nn\\
&&\threej{l_p}{\sf L}{l_n}{0}{0}{0}
\ninej{1\over2}{l_p}{j_p}{1\over2}{l_n}{j_n}{1}{\sf L}{\sf J},
\label{A8}\er
with the radial part $R_{\sf L}^0(pn;\kappa)$  given by \rf{A6}.

\vspace{0.5cm}
\subsection{Fermi function and effective momentum approximation (EMA)}

To account for the Coulomb interaction between  the charged lepton
and the residual nucleus, the QRAP code is setup to use  by default
the Fermi function~\cite{Bli66,Beh82}. This correction  was employed
in several works for reactions on $^{12}$C with neutrinos from the
DAR of $\mu^+$. As pointed out in Ref.~\cite{Vol00}, the quantity
$p_\ell R_A$ is of the order of 0.5, where $p_\ell$ is the lepton
momentum, and $R_A$ is the radius of the nucleus. Thus, the
correction is well described by a Fermi function. Yet, for high
energy neutrinos, e.g. neutrinos from the DIF of $\pi^+$, the
outgoing muons have $p_\ell R_A>0.5$. For these {\it relativistic}
leptons, the effective momentum approximation (EMA) \cite{Eng98}
should take care of the Coulomb field of the daughter nucleus,
instead of the Fermi function. This prescription for the Coulomb
correction is considered in the code within the subroutine
SECCION. More precisely,  with EMA=0 the Fermi function is
employed, while  with EMA=1 the EMA prescription is used. In the
EMA procedure, the lepton energy and momentum are modified by a
constant electrostatic potential within the nucleus
\brn
E_{\ell,eff}=E_\ell- V_{eff},&&
p_{\ell,eff}=\sqrt{E_{\ell,eff}^2 - m_{\ell}^2}, \ern
with
$V_{eff}=4V_C(0)/5=-6Z_f\alpha/5R_A$ \cite{Paa08,Ast07}. These
two approximations  for the Coulomb correction were tested in the
calculation of the inclusive cross section for neutrino scattering
on $^{208}$Pb \cite{Paa08}. As shown in Ref.\cite{Paa08}, the
Fermi function correction overestimates the cross sections
at higher neutrino energies where the EMA provides a more
reliable approach.  Thus, we recommend to use
the Fermi function correction in the range of neutrino energies
for which the cross section is below the corresponding EMA value,
whereas the EMA could be employed at higher energies, as shown in
previous studies \cite{Paa08,Vol00}.

As a final comment, the QRPA code could be easily extended to
calculate $\nu_\mu$-induced processes. This was done in
Refs.~\cite{Krm02,Krm05} to calculate $\nu_\mu-^{12}$C
cross sections using the EMA prescription for the DIF regime
of the LSND experiment.
The nuclear structure calculations remain the same, while the
kinematics changes by changing the electron mass to
the muon mass in the variable RMLEP.

%%%%%%%%%%%%%%%%%%%%%%%%%%%%%%%%%%%%%%%%%%%%%%%%%%%%%%%%%%%%%%%%%%%%%%%
%\newpage

\end{document}